\documentclass[reprint,article,superscriptaddress,msmath,amssymb,aps]{revtex4-1}
\usepackage{graphicx}
\usepackage{dcolumn}
\usepackage{bm}
\usepackage{color}
\usepackage[]{natbib}
\usepackage{amsmath,amssymb,amsfonts}

\begin{document}

\preprint{APS/123-QED}

\title{Radio-frequency capacitive gate-based sensing}


\author{Imtiaz Ahmed}\thanks{ia307@cam.ac.uk}
\affiliation{Cavendish Laboratory, University of Cambridge, J. J. Thomson Ave., Cambridge, CB3 0HE, United Kingdom}
\author{James A. Haigh}
\affiliation{Hitachi Cambridge Laboratory, J. J. Thomson Ave., Cambridge, CB3 0HE, United Kingdom}
\author{Simon Schaal}
\affiliation{London Centre for Nanotechnology, University College London, London, WC1H 0AH, United Kingdom}
\author{Sylvain Barraud}
\affiliation{CEA/LETI-MINATEC, CEA-Grenoble, 38000 Grenoble, France}
\author{Yi Zhu}
\affiliation{Department of Materials Science and Metallurgy, University of Cambridge, 27 Charles Babbage Road, Cambridge CB3 0FS, United Kingdom}
\author{Chang-min Lee}
\affiliation{Department of Materials Science and Metallurgy, University of Cambridge, 27 Charles Babbage Road, Cambridge CB3 0FS, United Kingdom}
\author{Mario Amado}
\affiliation{Department of Materials Science and Metallurgy, University of Cambridge, 27 Charles Babbage Road, Cambridge CB3 0FS, United Kingdom}
\author{Jason W. A. Robinson}
\affiliation{Department of Materials Science and Metallurgy, University of Cambridge, 27 Charles Babbage Road, Cambridge CB3 0FS, United Kingdom}
\author{Alessandro Rossi}
\affiliation{Cavendish Laboratory, University of Cambridge, J. J. Thomson Ave., Cambridge, CB3 0HE, United Kingdom}
\author{John J. L. Morton}
\affiliation{London Centre for Nanotechnology, University College London, London, WC1H 0AH, United Kingdom}
\affiliation{Department of Electronic \& Electrical Engineering, University College London, London WC1E 7JE, United Kingdom}
\author{M. Fernando Gonzalez-Zalba}\thanks{mg507@cam.ac.uk}
\affiliation{Hitachi Cambridge Laboratory, J. J. Thomson Ave., Cambridge, CB3 0HE, United Kingdom}

\begin{abstract}

Developing fast, accurate and scalable techniques for quantum state readout is an active area in semiconductor-based quantum computing. Here, we present results on dispersive sensing of silicon corner state quantum dots coupled to lumped-element electrical resonators via the gate. The gate capacitance of the quantum device is configured in parallel with a superconducting spiral inductor resulting in resonators with loaded Q-factors in the 400-800 range. For a resonator operating at 330~MHz, we achieve a charge sensitivity of 7.7~$\mu$e$/\sqrt{\text{Hz}}$ and, when operating at 616~MHz, we get 1.3~$\mu$e$/\sqrt{\text{Hz}}$. We perform a parametric study of the resonator to reveal its optimal operation points and perform a circuit analysis to determine the best resonator design. The results place gate-based sensing at par with the best reported radio-frequency single-electron transistor sensitivities while providing a fast and compact method for quantum state readout. 

\end{abstract}

\maketitle

  
\section{\label{sec:level2}Introduction}

The spins of isolated electrons in silicon are one of the most promising solid-state systems on which to implement quantum information processing. With the recent demonstrations of long coherence times~\cite{Veldhorst2014,kawakami2014electrical}, high fidelity spin readout~\cite{Morello2010}, and one- and two-qubit gates~\cite{Veldhorst2015,Kawakami2016,zajac2017resonantly,yoneda2017quantum,Watson2017}, the basic requirements to build a quantum computer have been fulfilled~\cite{DiVincenzo2000}. Now, scaling the technology to a number of qubits sufficiently large to perform computationally relevant calculations is one of the major objectives and several proposals for large scale integration have been put forward~\cite{Vandersypen2017,Veldhorst2017, Li2017}. In this respect, developing quantum state readout techniques that are fast and accurate while also being compact has become an active area of research. 

Conventionally, readout in semiconductor gate-defined qubits is achieved using sensitive external electrometers. The most prominent example is the single-electron transistor (SET). Its radio-frequency version, the rf-SET~\cite{schoelkopf1998radio}, sets the standard as the most sensitive electrometer with the best charge sensitivity reported to date ($0.9~\mu e/\sqrt{\text{Hz}}$~\cite{brenning2006ultrasensitive}). The enhanced performance is based on reflectometry techniques that use lumped-element LC circuits to match the high-resistance of the detector to the 50 $\Omega$ of the line~\cite{reilly2007fast}. However, mesoscopic electrometers, such as the rf-SET, need to be placed in close proximity to the qubits adding complexity to the circuit architecture. 

Circuit quantum electrodynamics (QED) offers an alternative method for state readout of a quantum system. In this case, the qubit is embedded in a high Q-factor on-chip microwave resonator. This can be to the point of strong coupling, where the qubit and microwave photon dynamics become hybridized. In the dispersive limit, when the resonator and the quantum system are detuned, the state of the qubit can then be directly inferred from the oscillatory state of the resonator. This has been used to read superconducting~\cite{Wallraff2004} and more recently, semiconductor qubits~\cite{Petersson2012,Stockklauser2017,mi2016strong}. 

The same principle of dispersive readout has also been applied to rf-reflectometry matching circuits~\cite{colless2013dispersive,gonzalez2015probing, house2015radio,Ares2016,crippa2017level}. This compact readout technique, namely in-situ gate-based readout, uses existing gate electrodes coupled to off-chip lumped-element resonators for sensing~\cite{colless2013dispersive,gonzalez2015probing}. This method alleviates the burden of external electrometers and reduces the complexity of the qubit architecture. Typically, gate-based sensing has been performed using low Q-factor resonators inspired by the matching networks developed for rf-SETs~\cite{colless2013dispersive,gonzalez2015probing, house2015radio,Ares2016,crippa2017level} and have not been optimized for reactive changes in device characteristics, such as the quantum or tunnelling capacitance~\cite{Ashoori1992, ciccarelli2011impedance,persson2010excess,Mizuta2017}.

Here, we bridge the gap between circuit QED type measurements and lumped-element reflectometry techniques, by optimizing external matching circuits for capacitive changes. We show that significant improvements in sensitivity are possible by changing the circuit topology to enhance the Q-factor of the resonator. While in one sense this brings rf-reflectometry towards conventional circuit QED, the fact that we keep the microwave circuitry separate means that it can be fabricated separately from the nanodevice. This allows independent nanofabrication strategies for the resonator and the qubit. In particular, devices can be optimized to have a large coupling to the quantum system which is an important ingredient for sensitive dispersive readout. This is were complementary metal-oxide-semiconductor (CMOS) technology, that for decades has been optimising the gate coupling to the channel, can give an advantage over other qubit platforms. 

In this paper, we report on in-situ dispersive readout of silicon-based CMOS few-electron quantum dots (QDs)~\cite{betz2014high, voisin2014few} using lumped-element classical resonators with loaded quality factors, $Q_\text{L}$, in the 400-800 range. The enhancement in $Q_\text{L}$ is achieved by configuring the device gate capacitance in parallel with a superconducting spiral inductor and coupling via a coupling capacitor to a PCB coplanar waveguide. We find charge sensitivities of $7.7~\mu e/\sqrt{\text{Hz}}$ and $1.3~\mu e/\sqrt{\text{Hz}}$ for resonators operating at 330~MHz and 616~MHz, respectively. The latter represents an improvement of a factor of 30 over previous gate-based sensors~\cite{gonzalez2015probing} and sets gate-based reflectometry at par with the best ever reported rf-SETs. Finally, following a circuit analysis, we summarize the key requirements for sensitive capacitive gate-based readout: large $Q_\text{L}$ resonators, well matched to the line, with low parasitic capacitance and large gate coupling to the quantum system. Our results pave the way for time-resolved dispersive readout of electron spin dynamics.

%


\section{\label{sec:level2} Device and Resonator}

\begin{figure}
	\begin{center}
		\includegraphics[scale=1]{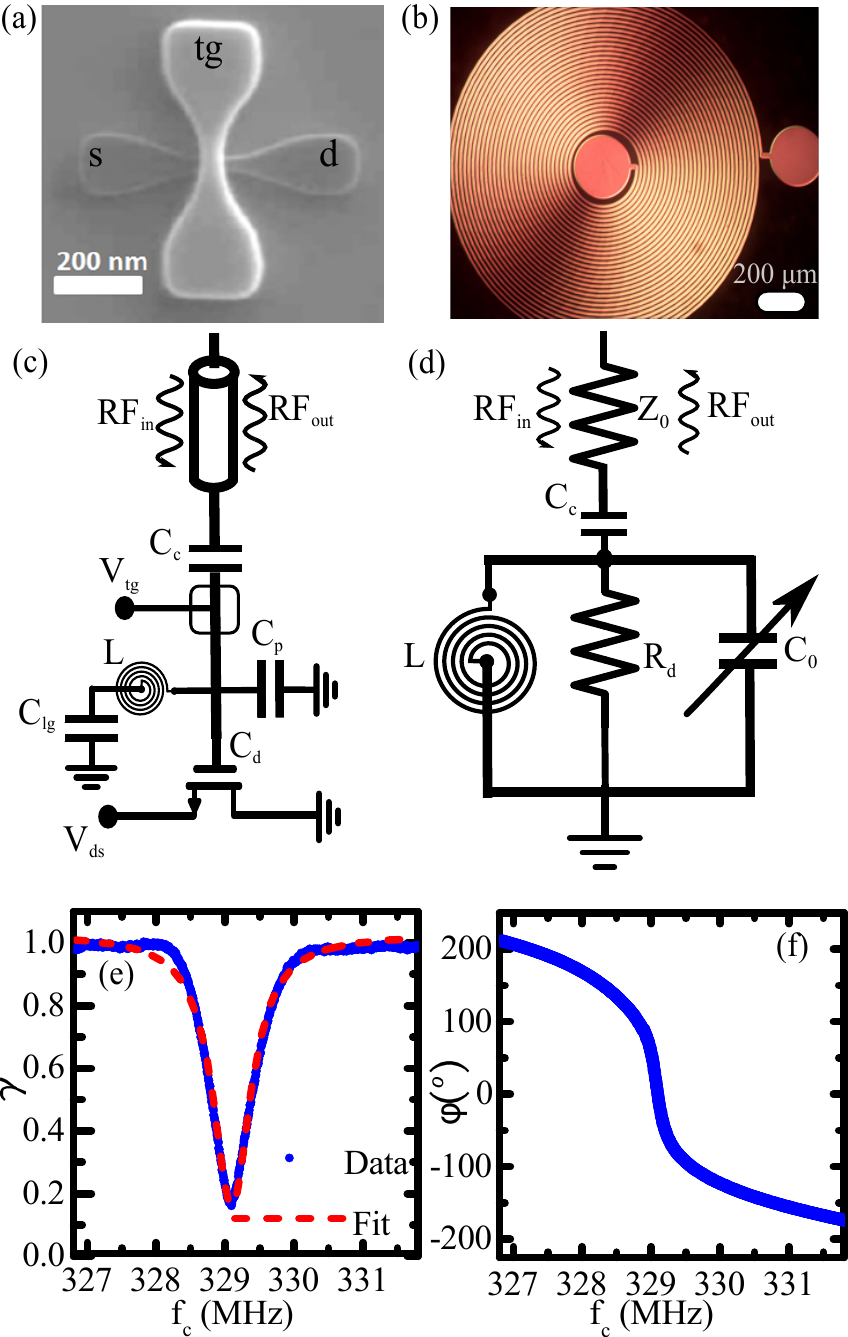}
	\end{center}
	\caption{ Device and resonator. (a) A scanning-electron micrograph of a NWFET showing source (s), drain (d) and top-gate (tg) terminals. (b) Optical image of a superconducting NbN spiral inductor. The spiral track width and track spacing are both 8~$\mu$m. (c) Circuit diagram for rf-reflectometry. The NbN inductor $L$ is connected in parallel with the top-gate of the NWFET. The circuit has a parasitic capacitance of $C_\text{p}$ to ground. $V_\text{ds}$ and $V_\text{tg}$ are the bias voltages. $C_\text{lg}=100$~pF. (d) Model for the parallel resonator coupled to external line impedance $Z_0=50$~$\Omega$ through $C_\text{c}$. The resistor $R_\text{d}$ represents the losses in the resonator. (e) Amplitude $\gamma$ of the reflection coefficient $\Gamma$ measured (blue) and fit (red). (f) Phase $\phi$ of the reflection coefficient $\Gamma$.}
	\label{Figure_01}
\end{figure}

The device investigated is a CMOS silicon nanowire field-effect transistors (NWFET) with channel length $l=30$~nm, width $w=60$~nm and height $h=11$~nm similar to the one shown in Fig.~\ref{Figure_01}(a). The top-gate (tg) wraps around three faces of the n-type channel between the highly-doped source (s) and drain (d). At low temperatures, when the NWFET is biased below threshold ($V_\text{tg} \approx 0.5V $), few-electron QDs form in the NW channel~\cite{sellier2007subthreshold,betz2014high}. The transistor's multigate geometry, combined with a small equivalent gate oxide thickness of 1.3~nm, results in QDs with large gate couplings $\alpha=C_\text{tg}/C_\Sigma=0.85-0.89$ since the total capacitance $C_\Sigma$ is mostly given by its capacitance to the gate electrode $C_\text{tg}$. 

The device is embedded in an electrical resonator containing a polycrystalline NbN superconducting planar spiral which provides a low-loss and low self-capacitance inductor $L$~\cite{hornibrook2014frequency,stevenson2002multiplexing,colless2013dispersive}. The 80~nm  NbN films were grown in unheated c-plane 430~$\mu$m thick sapphire substrates by DC magnetron sputtering. The deposition was done in an Ar/N$_2$ atmosphere with 28\% N$_2$ at 1.5~Pa. The spiral was defined using optical lithography and etching, see Fig.~\ref{Figure_01}(b). We wirebonded the inductor in parallel to the device gate capacitance to ground, $C_\text{d}$, and the circuit parasitic capacitance, $C_\text{p}$, and coupled to the $Z_0$ line via a coupling capacitor $C_\text{c}$ as shown in Fig.~\ref{Figure_01}(c). This differs from the series configuration explored in Refs.~\cite{colless2013dispersive,gonzalez2015probing,house2015radio,Ares2016,crippa2017level} and, as we shall see later, leads to enhanced sensitivity to capacitance changes. A simple equivalent model for the resonator, as in Fig.~\ref{Figure_01}(d), consists of $L$, the circuit losses $R_\text{d}$ and the variable capacitor $C_0=C_\text{p}+C_\text{d}$ placed in parallel and coupled to the line by $C_\text{c}$. $R_\text{d}$ represents dielectric losses in the device and the PCB, and can contain dissipative terms arising from Sisyphus processes~\cite{persson2010excess,ciccarelli2011impedance,Lambert2014,gonzalez2015probing}. The parasitic capacitance, $C_\text{p}$, combines contributions from the device and the PCB.  

In order to characterize the resonant frequency $f_0$, bandwidth BW and $Q_\text{L}$, we measure the complex reflection coefficient $\Gamma = \gamma e^{i \phi}$ as a function of the carrier frequency $f_\text{c}$. In Fig.~\ref{Figure_01}(e), we show the magnitude $\gamma$ (in blue) and a fit (in red). From this we estimate $L=405$~nH, $C_\text{c}$ = 90~fF, $C_\text{p}$ = 480~fF, $R_\text{d}$ = 800 k$\Omega$. This gives us $f_{0} = 1/(2\pi\sqrt{L(C_\text{c} +C_\text{0})}) = 329.33$ MHz, $Q_\text{L}\approx 400$ and BW$ = 0.82$ MHz. The large depth of the resonance, $\gamma_ {min} = 0.168$ indicates that the resonator is close to matching. The loaded Q contains contributions from the external Q-factor, $Q_\text{e}=(C_\text{c}+C_\text{0})/2\pi f_0Z_0C_\text{c}^2=680$, and the unloaded Q-factor of the resonator, $Q_0 = 2\pi f_0(C_\text{c}+C_\text{0})R_\text{d}=943$. In this particular design, external losses dominate $Q_\text{L}$ but its value is increased by an order of magnitude when compared to series resonator gate-based approaches~\cite{colless2013dispersive,gonzalez2015probing,house2015radio,Ares2016,crippa2017level}. We operate in the overcoupled regime confirmed by the 180$^o$ phase shift, $\phi$, as a function of carrier frequency in Fig.~\ref{Figure_01} (e). 



\section{\label{sec:level2} Dispersive Regime}

Gate-based sensing is a resonant technique that allows probing the complex admittance of a quantum device~\cite{colless2013dispersive,chorley2012measuring,ciccarelli2011impedance,petersson2010charge}. Here, we couple a single QD in the NW channel to the resonator and probe its impedance using gate-based radio-frequency reflectometry at 40~mK~\cite{gonzalez2016gate,urdampilleta2015charge}. We use this to probe susceptance changes when adiabatic single-electron tunneling occurs between the QD and the s/d reservoirs. At resonance, variations in $\phi$ capture changes in the device capacitance $\Delta C$ that can be attributed to tunneling or quantum capacitance~\cite{Mizuta2017}. Since the resonator is overcoupled, $\Delta\phi=-2Q_\text{L}\Delta C/(C_\text{c} +C_\text{0})$. 

\begin{figure}
\begin{center}
\includegraphics[scale=1]{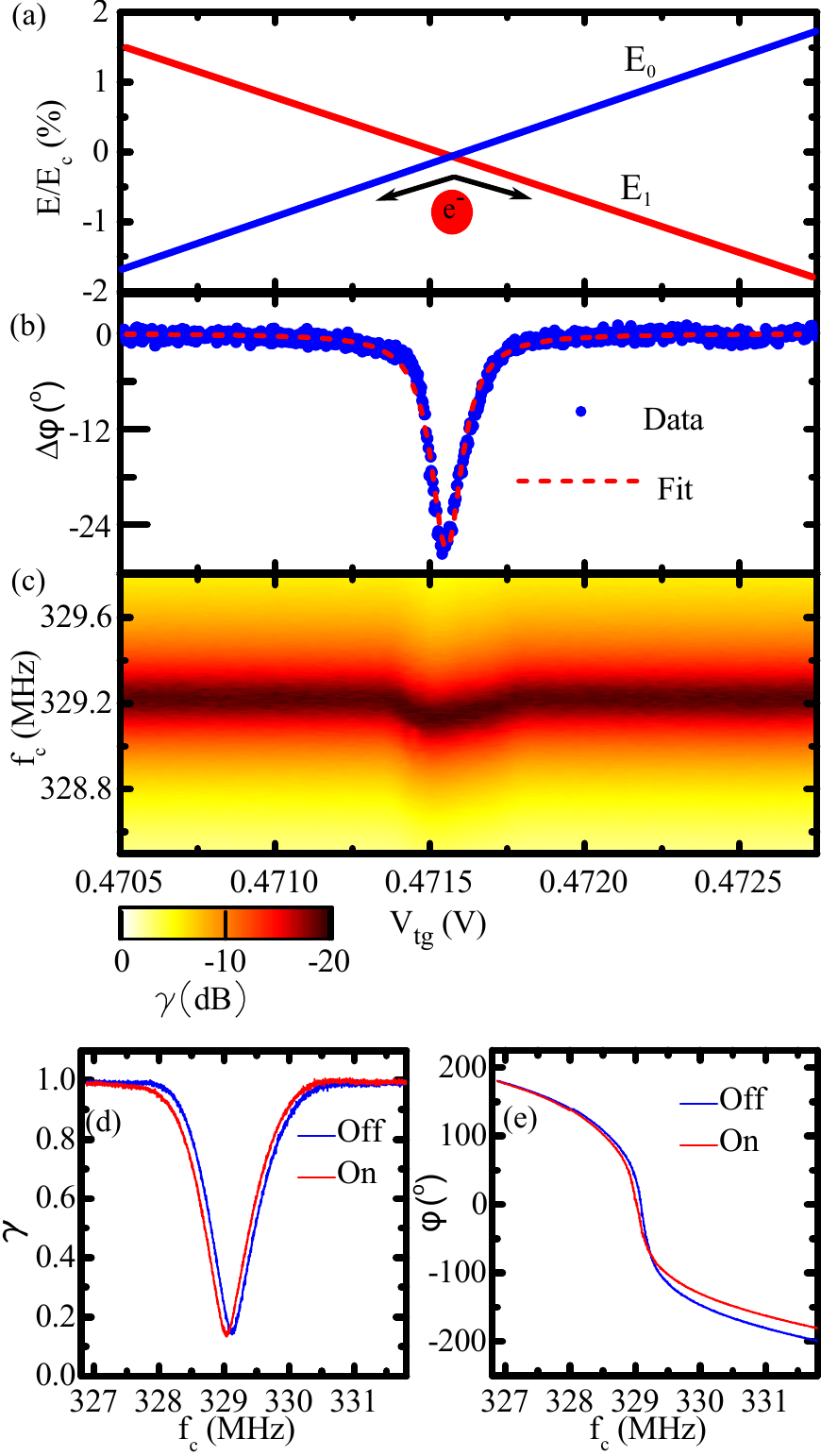}
\end{center}
\caption{ Dispersive regime. (a) Energy diagram of a fast driven TLS, $E_0$ and $E_1$, across a charge degeneracy point. The energy is normalized to charging energy $E_\text{c}$ of the QD. (b) Data (blue dot) and Lorentz fit (red dashed line) for the phase change $\Delta\phi$ of the resonator as a function of $V_\text{tg}$. (c) $\gamma$ as function of carrier frequency $f_\text{c}$ and $V_\text{tg}$. Experimental data for $\gamma$ (d)  and $\phi$ (e) at two different $V_\text{tg}$ voltages - away from and at the charge degeneracy, blue and red traces respectively.}
\label{Figure_02}
\end{figure}

In our system, the origin of $\Delta C$ can be explained by considering an uncoupled two level system (TLS) described by a QD with zero ($E_0$) or one excess electron ($E_1$). Particles are exchanged with the s/d reservoirs. If the TLS is driven by an external sinusoidal excitation at $f_{0}$ and the relaxation rate $\nu$ between levels is comparable, Sysiphus dissipation occurs~\cite{persson2010excess,gonzalez2015probing}. However, if $\nu\gg f_{0}$, electrons tunnel adiabatically and out of phase with the drive. This results in a purely dispersive signal manifesting as a tunneling capacitance contribution $C_\text{t}$~\cite{house2015radio}. In Fig.~\ref{Figure_02}(a), we show schematically a TLS driven across a charge degeneracy point where $E_0$ and $E_1$ cross each other at a fixed top-gate voltage point $V_\text{tg}^0$. In the regime $\nu \gg f_{0}$, the electron always stays in the ground state and the tunneling capacitance is given by 


\begin{equation}\label{tunel}
C_\text{t} = \frac{(\alpha e)^2}{\pi}\frac{h\nu}{(h\nu)^2 + (\alpha e \Delta V_\text{tg})^2},
\end{equation}

\noindent where $h$ is the Planck's constant, $e$ is the electron charge and $\Delta V_\text{tg}=V_\text{tg}-V_\text{tg}^0$~\cite{house2015radio,cottet2011mesoscopic}. This voltage-dependent $C_\text{t}$ produces a phase shift in the resonator as can be seen in Fig.2 (b) where $\Delta \phi$ is plotted as the $V_\text{tg}$ is swept across the charge degeneracy point. We measure a maximum phase shift $\Delta \phi = 28^o$ at the degeneracy point. Using Eq.~(\ref{tunel}), we fit $\Delta \phi$ (red curve) and extract $\nu=26$~GHz ($\gg f_0$) from the full-width-half-maximum (FWHM). $C_\text{t}$ loads the resonator pulling down its resonant frequency as shown in Fig.~\ref{Figure_02}(c). Finally, in Fig.~\ref{Figure_02}(d,e), we compare $\gamma$ and $\phi$ on and off the degeneracy point. From these measurements we extract a dispersive shift $\Delta f=88$~kHz, which corresponds to an effective change in capacitance given by $\Delta C=2(C_\text{c} +C_\text{0})\Delta f/f_0=0.3$~fF. This agrees well with the expected maximum tunneling capacitance 0.37~fF, calculated from Eq.~(\ref{tunel}).


\section{\label{sec:level2} Charge Sensitivity}

\begin{figure}
\begin{center}
\includegraphics[scale=1]{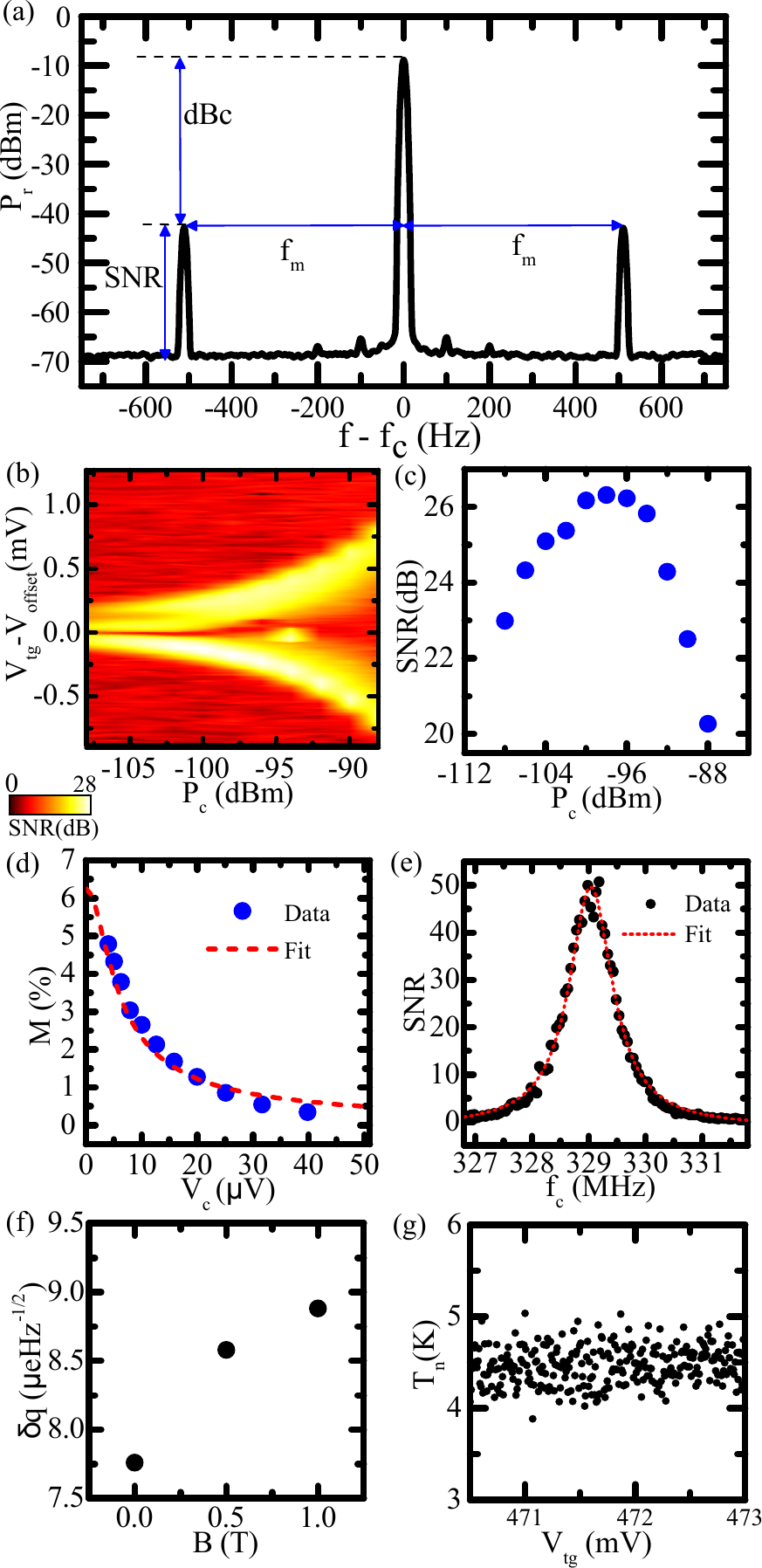}
\end{center}
\caption{Measuring and optimizing charge sensitivity. (a) Reflected power $P_\text{r}$ spectrum showing the sidebands at $f_\text{c}\pm f_\text{m}$ due to gate-voltage modulation. (b) SNR in dB as a function of $V_\text{tg}$ and $P_\text{c}$ ($V_\text{offset}=472.23$~mV). (c)~Sideband SNR as function of $P_\text{c}$. (d) Modulation index $M$ (blue dots) and fit (red dashed line) as a function of input rf carrier voltage $V_\text{c}$. (e) Sideband power SNR in linear scale vs. rf-carrier frequency $f_\text{c}$ measured at $P_\text{c}=-115$~dBm. (f)~Measured charge sensitivity as a function of the magnetic field $B$. (g) Noise temperature $T_\text{n}$ as a function of $V_\text{tg}$ measured at $f_0$ with RBW=300~kHz.}
\label{Figure_03}
\end{figure}

We use the conventional technique~\cite{schoelkopf1998radio} to measure the charge sensitivity of the capacitive gate-based sensor: A small sinusoidal voltage is applied to the top-gate of the device with a root mean square charge equivalent amplitude $\Delta q$ and frequency $f_\text{m}$. This produces an amplitude modulation of the carrier that results in sidebands appearing in the power spectrum of the reflected signal at $f_\text{c}\pm f_\text{m}$, as can be seen in Fig.~\ref{Figure_03}(a) along with the rf carrier. The height of the sideband measured from the noise floor defines the power signal-to-noise ratio (SNR). The charge sensitivity is then calculated from the definition, $\delta q = \Delta q/(\sqrt{2\text{RBW}}\times 10^{\frac{\text{SNR}}{20}})$ where RBW is the resolution bandwidth of the spectrum analyzer~\cite{brenning2006ultrasensitive}. A separate figure of merit is the modulation depth dBc, given by the sideband height relative to the carrier in dB. This figure indicates how much of the input signal is modulated by the device.

Fig.~\ref{Figure_03}(a) shows the power spectrum at the optimal working point of the resonator (identified by measurements described below). The spectrum is obtained using a modulation signal with $f_\text{m}=511$~Hz and $\Delta q=6.98\times 10^{-3}e$ and a $\text{RBW}=10$~Hz. We measure a SNR=26.1~dB, resulting in $\delta q=7.7$~$\mu$e/$\sqrt{\text{Hz}}$. This result represents a charge sensitivity improvement of a factor of $\backsim 5$ with respect to previous reports~\cite{gonzalez2015probing} and demonstrates the advantage of adopting the circuit configuration of the capacitive gate sensor. 

We now discuss the parametric study of the gate-based sensor's sensitivity in terms of $V_\text{tg}$, $f_\text{c}$ and carrier power $P_\text{c}$ to find the optimal working point. In Fig.~\ref{Figure_03}(b), we show the lower sideband SNR as a function of $V_\text{tg}$ and $P_\text{c}$. For a fixed $P_\text{c}$, the SNR shows two maxima whose separation in $V_\text{tg}$ increases as $P_\text{c}$ is increased. The position of the maxima corresponds to the $V_\text{tg}$ points of maximum slope at either side of the charge transition in Fig.~\ref{Figure_02}(b). The dependence of the separation with increasing $P_\text{c}$ indicates that the transition is being broadened by the rf voltage $V_\text{c}$. Only at the lowest values of $P_\text{c}$, where the separation between peaks remains constant, is the transition lifetime limited. 

To find the optimal carrier power for sensing, we extract the maximum SNR at each $P_\text{c}$ (blue dots) as shown in Fig.~\ref{Figure_03}(c). The SNR peaks at $P_\text{c}=-98$~dBm when the transition is still power broadened, as opposed to the expectation that the maximum SNR would be achieved when the transition is lifetime broadened. To understand why, we define the modulation index $M=10^\frac{\text{dBc}}{20}$ and note that the sideband SNR can be expressed as SNR$=M\times P_\text{c}$. $M$ decreases as the rf carrier voltage $V_\text{c}$ increases as seen in Fig.~\ref{Figure_03}(d). This dependence can be modeled as a convolution of two competing processes: lifetime broadening and power broadening. The transition is lifetime broadened by $\nu$ to produce a linewidth $V_{\nu}$ ($h\nu = e\alpha V_{\nu}$ ) and power broadened by applied rf carrier power with a linewidth proportional to $V_\text{c}$. The dependence of $M$ on $V_\text{c}$ can be approximated by $ 1/\sqrt{V_\nu^2 + V_{c}^2}$ which decreases with increasing $V_\text{c}$ as shown in Fig.~\ref{Figure_03}(e) (red dashed curve). The maximum SNR occurs when the increase in input power is compensated by the decrease in $M$.

Next, we find the optimal carrier frequency. We plot the sideband power SNR (in linear scale) at $f_{c}-f_m$ as a function of rf carrier frequency $f_\text{c}$ when swept across a frequency range containing $f_0$ (see Fig.~\ref{Figure_03}(e)). The SNR shows a Lorentzian profile with center frequency $f_0$, and BW and $Q_\text{L}$ matching the values obtained from Fig.~\ref{Figure_01}(e). To avoid power broadening, the input RF power was kept at $-115$~dBm during this measurement.


Additionally, we study the dependence of the sensor's charge sensitivity on in-plane magnetic field, given our use of a superconducting material (NbN) for the lumped element inductor, and the fact that for typical spin qubit systems, an external magnetic field is used to Zeeman-split the spin degenerate energy levels~\cite{hanson2007spins,Morello2010,Veldhorst2015}. We measured SNR$=24.9$~dB at 1~T which gives a charge sensitivity $\delta q =8.8$~$\mu$e$/\sqrt{\text{Hz}}$ as shown in Fig.~\ref{Figure_03}(f). This result demonstrates that our gate-sensor sensitivity only deteriorates by 15\% at 1~T and hence is robust against moderate magnetic fields used to operate Si spin qubits.

Finally, in Fig.~\ref{Figure_03}(g), we measure the noise temperature $T_\text{n}$ of the system at the resonance frequency as a function of ${V}_\text{tg}$. As we sweep ${V}_\text{tg}$ across the charge degeneracy point, $T_\text{n}$ stays constant $4.5\pm 0.4$~K which matches with the noise temperature of our cryogenic amplifier (Quinstar QCA-U350-30H). Hence we conclude that charge sensitivity is limited by the thermal noise of the cryo-amplifier and not by Sisyphus noise which can be orders of magnitude smaller~\cite{gonzalez2015probing}. In this study, we emphasize the importance of improving the resonator design to increase the signal of gate-based approaches. Improving the noise floor by using, for example, a Josephson Parameter Amplifier (JPA) will lead to additional enhancements on the experimental sensitivity of the capacitive gate-based sensor.

\section{\label{sec:theory} Resonator Optimization}

In this section, we explore the resonator design analytically to highlight ways to optimize the circuit and understand the ultimate performance of capacitive gate-based charge sensing. We consider the circuit in Fig.~\ref{Figure_01}(d) and its reflection coefficient:

\begin{equation}
	\Gamma=\frac{Z-Z_0}{Z+Z_0},
\end{equation}

\noindent where Z is the complex impedance of the coupling capacitor $C_\text{c}$ in series with the parallel combination of the inductor $L$, the circuit resistance $R_\text{d}$ and circuit variable capacitance $C_\text{0}$. Gate-based reflectometry is sensitive to changes in the reflection coefficient. In this work, we are concerned with capacitance changes in the circuit due to single-electron tunneling that can manifest in the form of quantum or tunneling capacitance~\cite{Mizuta2017}. Therefore, we calculate the absolute value of the differential change in reflection coefficient with $C_0$,

\begin{equation}\label{Gammaderiv}
	\left|\Delta\Gamma\right|=\left|\frac{\partial\Gamma}{\partial C_\text{0}} \Delta C\right|=\frac{2R_\text{eq}Z_0}{(R_\text{eq}+Z_0)^2}Q_0\frac{\Delta C}{C_\text{c}+C_\text{0}}
\end{equation}

\begin{figure}
	\centering
		\includegraphics{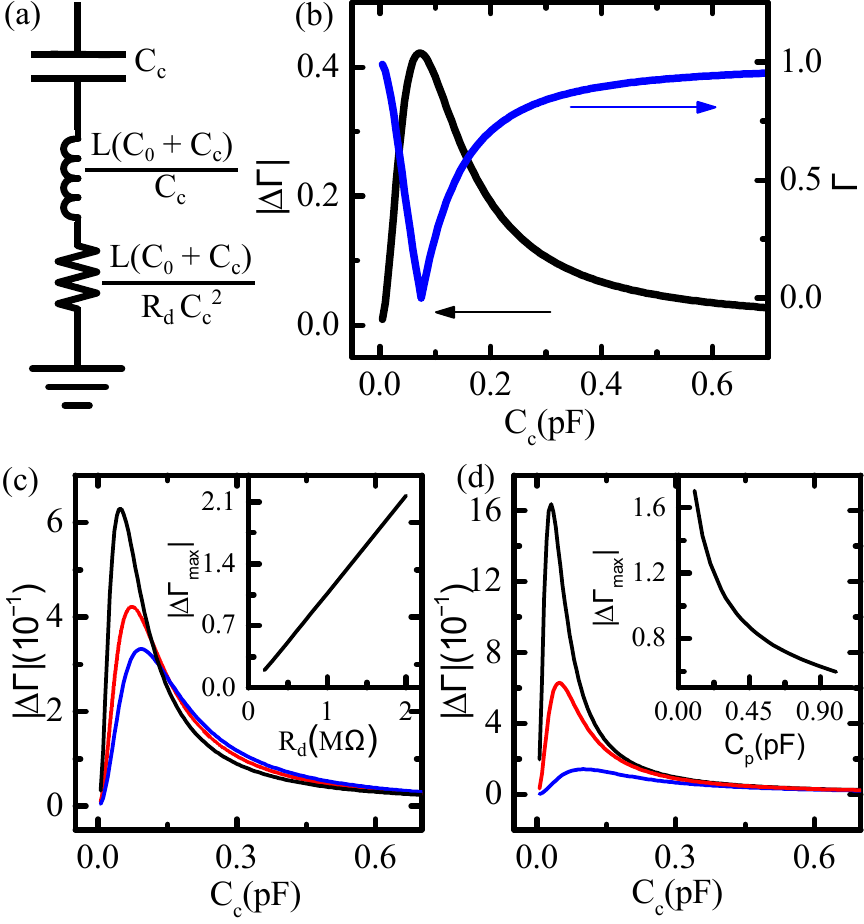}
	\caption{Resonator optimization. (a) Equivalent circuit at resonance. (b) $\left|\Delta\Gamma\right|$ (black) and $\Gamma$ (blue) as a function of $C_\text{c}$ for the experimental values $R_\text{d}=800$~k$\Omega$, $C_\text{p}= 0.48$~pF and $L=405$~nH, $Z_0=50$~$\Omega$. We consider $\Delta C=1$~fF. (c) Dependence on $R_\text{d}$. $\left|\Delta\Gamma\right|$ as a function of $C_\text{c}$ for $R_\text{d}=200$~k$\Omega$ (blue), 800~k$\Omega$ (red) and 2~M$\Omega$ (black). Inset: Maximum $\left|\Delta\Gamma\right|$ as a function of $R_\text{d}$. (d) (c) Dependence on $C_\text{p}$. $\left|\Delta\Gamma\right|$ as a function of $C_\text{c}$ for $C_\text{p}=0.8$~pF (blue), 0.48~pF (red) and 0.2~pF (black). Inset: Maximum $\left|\Delta\Gamma\right|$ as a function of $C_\text{p}$.}
	\label{Figure_04}
\end{figure}

Here, $R_\text{eq}=L(C_\text{c}+C_\text{0})/R_\text{d}C_\text{c}^2$, is the equivalent resistance of the circuit at $f=f_0$, as depicted in Fig.~\ref{Figure_04}(a). The charge sensitivity is inversely proportional to $\left|\Delta\Gamma\right|$ and hence a study of this magnitude yields an estimate of the relative sensitivity level~\cite{roschier2004noise,muller2013circuit}.

Eq.~\ref{Gammaderiv} provides the guidelines to optimize the sensitivity of gate-based sensing approaches. Firstly, exemplified in Fig.~\ref{Figure_04}(b), where we plot both $\left|\Delta\Gamma\right|$ and $\Gamma$ as a function of $C_\text{c}$, we observe that $\left|\Delta\Gamma\right|$ is maximum when the coupling capacitor is chosen to give perfect matching ($\Gamma=0$) i.e. $R_\text{eq}=Z_0$. Secondly, increasing $R_\text{d}$, and in turn the unloaded Q-factor, leads to an increase in sensitivity. The effect on $\left|\Delta\Gamma\right|$ of increasing $R_\text{d}$ can be seen in Fig.~\ref{Figure_04}(c) where the maximum $\left|\Delta\Gamma\right|$ increases as $R_\text{d}$ is increased from 200~k$\Omega$ (blue) to 2~M$\Omega$ (black). Note the change in optimal $C_\text{c}$ as $R_\text{d}$ is varied. $\left|\Delta\Gamma\right|_\text{max}$ increases linearly with $R_\text{d}$ as can be seen in the inset. Thirdly, a reduction of the circuit capacitance, by reducing $C_\text{p}$, leads to an enhanced sensitivity. This can be observed in Fig.~\ref{Figure_04}(c) where we plot $\left|\Delta\Gamma\right|$ as a function of $C_\text{c}$ and fixed $L$ and $R_\text{d}$ for three different values of $C_\text{p}$, 0.2~pF (black), 0.48~pF (red) and 0.8~pF (blue). $\left|\Delta\Gamma\right|_\text{max}$ decreases as $C_\text{p}^{-1/2}$, as can be seen in the inset. Finally, the change in device capacitance, $\Delta C$, needs to be maximized. This can be achieved by maximising the gate coupling factor $\alpha$ which has a quadratic effect on $\left|\Delta\Gamma\right|$ as can be seen in Eq.~(\ref{tunel}).

Ultimately, Eq.~\ref{Gammaderiv} can be expressed in much simpler terms when the resonator is matched to the line, $\left|\Delta\Gamma\right|=\pi R_\text{d} f_0\Delta C$. In this case, we see that the optimal device should have as low dissipation as possible (large $R_\text{d}$). Moreover, gate-based sensing benefits from operating at high frequency as we demonstarte in the next Section. 

\section{Higher frequency operation}

\begin{figure}
	\centering
		\includegraphics{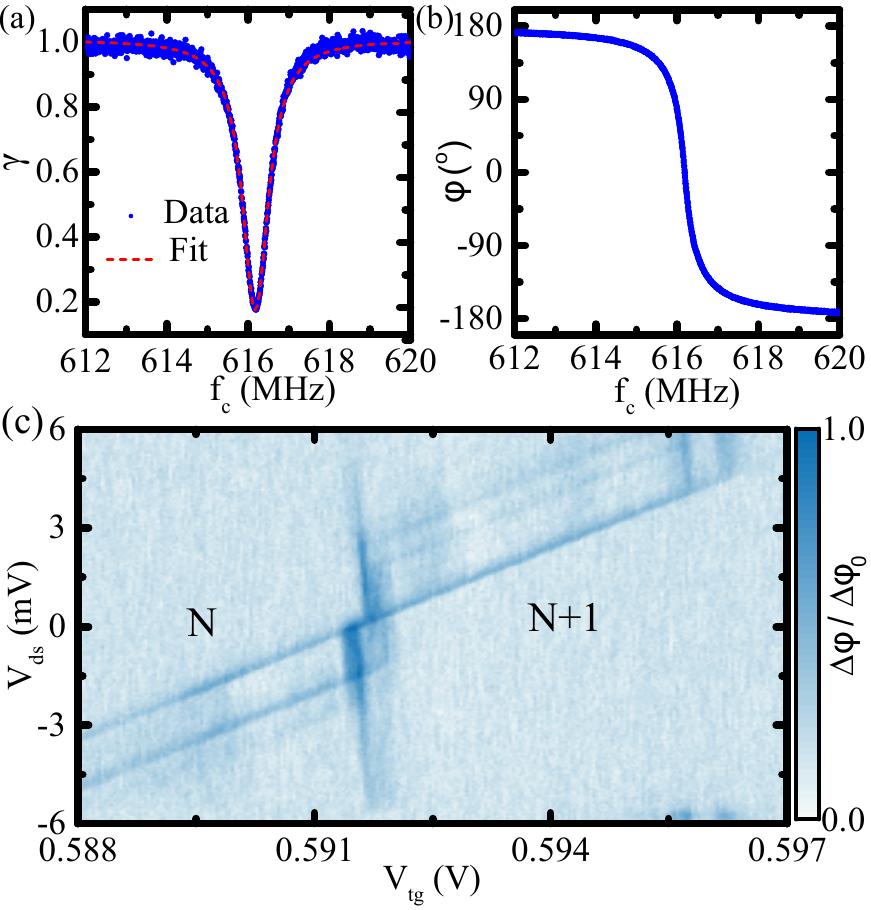}
	\caption{High frequency resonator. Magnitude (a) and phase (b) of the reflection coefficient $\Gamma$ as a function of frequency. (c) Fast data acquisition of a dot to reservoir transition. $V_\text{ds}$ was ramped at 4~kHz while $V_\text{tg}$ at 7~Hz. Each trace was averaged 5 times. The total measurement time is 700~ms.}
	\label{Figure_05}
\end{figure}

To asses the advantage of operating the capacitive gate-based sensor at higher frequencies, we perform a second set of experiments on a nominally identical device but narrower channel, $w=30$~nm, and a resonator with resonant frequency $f_0=616.18$~MHz. We use a NbN inductor, $L=134$~nH. In Fig.~\ref{Figure_05}(a,b), we see the magnitude $\gamma$ and phase $\phi$ of the reflection coefficient. The resonator has a $BW=0.78$~MHz and hence a loaded Q-factor $Q_\text{L}=790$, is close to matching $\gamma_\text{min}=0.1$ and overcoupled $Q_\text{e}<Q_0$.

We characterize the gate-based sensor in terms of charge sensitivity following the procedure explained in Section IV and obtain an optimized charge sensitivity of $1.3~\mu$e$/\sqrt{\text{Hz}}$ at $P_\text{c}=-120$~dBm. For this measurement, we use $\Delta q=1.48\times 10^{-4}e$ and $\text{RBW}=20$~Hz. The sensitivity improvement is $\approx 30$ when compared to previous reports~\cite{gonzalez2015probing} and places the charge sensitivity of the capacitive gate-based sensors on a par with the best reported rf-SET sensitivities~\cite{aassime2001radio,brenning2006ultrasensitive}.

We demonstrate the advantage of the improved sensitivity by acquiring a charge stability map of the device containing $512 \times 256$ data points in just 700~ms, see Fig.~\ref{Figure_05}(c). Here, we use a double ramp scheme~\cite{StehlikPRapp2015} where we ramp the drain voltage, $V_\text{ds}$ at 4~kHz (sawtooth) while slowly ramp $V_\text{tg}$ at 7~Hz (triangular). The frequency of the ramp is limited in this measurement by the low-pass filtering in our lines (cut-off frequency 5~kHz). The 2D map is composed of traces of the demodulated phase response. In Fig.~\ref{Figure_05}(c), we see the characteristic signature of Coulomb Blockade measured dispersively with the gate-based sensor. The combination of the data quality and short acquisition time demonstrates the potential of this new gate-based sensor design for fast readout of semiconductor nanostructures. Given the integration time of 5~$\mu$s per point, capacitive gate-based sensing may enable performing single-shot readout of electron spin dynamics in silicon. 

\section{\label{sec:level2} Conclusion}

We have demonstrated an optimized design for in-situ gate-based sensing for which the charge sensitivity is a factor of 28 better than the best reported gate-sensor~\cite{gonzalez2015probing} and it is comparable to the best sensitivities ever demonstrated for the RF-SET~\cite{brenning2006ultrasensitive}. In the case of the RF-SET the experimental sensitivity is limited by shot-noise whereas in the case of our gate-sensor is limited by the noise of the cryogenic amplifier. Hence by using a quantum-limited amplifier, a gate-sensors with sub $\mu$e$/\sqrt{\text{Hz}}$ sensitivity will be possible, making the gate-sensor the most sensitive electrometer. The ultimate sensitivity of this dispersive sensor remains to be explored and a study should consider the effects of the Sysiphus noise~\cite{gonzalez2015probing} and the Johnson-Nyquist noise of the resonator which can be orders of magnitude lower than shot-noise at miliKelvin temperatures and radio-frequencies. Additionally, the charge sensitivity demonstrated combined with the bandwidth of the resonators results in a few microsecond detection with an rms charge noise ($\Delta Q=\delta q\times\sqrt{\text{BW}}$) well below $1~e$. Therefore, capacitive gate-based sensing could allow single-shot readout of electron spins in silicon double QDs where the relaxation and coherence times are typically larger than a microsecond~\cite{Veldhorst2014,kawakami2014electrical}. In the future, devices with additional gates such as CMOS transistors in series~\cite{maurand2016cmos} or split-gate CMOS transistors~\cite{betz2015dispersively,betz2016reconfigurable} should provide access to experiments in which the resonator couples to an interdot charge transitions and Pauli spin blockade is used for electron spin readout.

\section{Acknowledgments}

We thank Ferdinand Kuemmeth and Andrew J. Ferguson for useful comments. This research has received funding from the European Union's Horizon 2020 Research and Innovation Programme under grant agreement No 688539 (http://mos-quito.eu) and No. 732894 (http://hot-fetpro.eu); as well as by the Engineering and Physical Sciences Research Council (EPSRC) through the Centre for Doctoral Training in Delivering Quantum Technologies (EP/L015242/1) and UNDEDD (EP/K025945/1), and the Winton Programme of the Physics of Sustainability. I.A. is supported by the Cambridge Trust and the Islamic Development Bank. A. R. acknowledges support from the European Union's Horizon 2020 research and innovation programme under the Marie Sklodowska-Curie grant agreement No 654712 (SINHOPSI).  J.W.A.R. acknowledges funding from the Royal Society and M.A. the EPSRC through the Programme Grant EP/N017242/1 and EP/P026311/1.

\bibliographystyle{apsrev4-1}
\bibliography{CS}

\begin{thebibliography}{47}%
\makeatletter
\providecommand \@ifxundefined [1]{%
 \@ifx{#1\undefined}
}%
\providecommand \@ifnum [1]{%
 \ifnum #1\expandafter \@firstoftwo
 \else \expandafter \@secondoftwo
 \fi
}%
\providecommand \@ifx [1]{%
 \ifx #1\expandafter \@firstoftwo
 \else \expandafter \@secondoftwo
 \fi
}%
\providecommand \natexlab [1]{#1}%
\providecommand \enquote  [1]{``#1''}%
\providecommand \bibnamefont  [1]{#1}%
\providecommand \bibfnamefont [1]{#1}%
\providecommand \citenamefont [1]{#1}%
\providecommand \href@noop [0]{\@secondoftwo}%
\providecommand \href [0]{\begingroup \@sanitize@url \@href}%
\providecommand \@href[1]{\@@startlink{#1}\@@href}%
\providecommand \@@href[1]{\endgroup#1\@@endlink}%
\providecommand \@sanitize@url [0]{\catcode `\\12\catcode `\$12\catcode
  `\&12\catcode `\#12\catcode `\^12\catcode `\_12\catcode `\%12\relax}%
\providecommand \@@startlink[1]{}%
\providecommand \@@endlink[0]{}%
\providecommand \url  [0]{\begingroup\@sanitize@url \@url }%
\providecommand \@url [1]{\endgroup\@href {#1}{\urlprefix }}%
\providecommand \urlprefix  [0]{URL }%
\providecommand \Eprint [0]{\href }%
\providecommand \doibase [0]{http://dx.doi.org/}%
\providecommand \selectlanguage [0]{\@gobble}%
\providecommand \bibinfo  [0]{\@secondoftwo}%
\providecommand \bibfield  [0]{\@secondoftwo}%
\providecommand \translation [1]{[#1]}%
\providecommand \BibitemOpen [0]{}%
\providecommand \bibitemStop [0]{}%
\providecommand \bibitemNoStop [0]{.\EOS\space}%
\providecommand \EOS [0]{\spacefactor3000\relax}%
\providecommand \BibitemShut  [1]{\csname bibitem#1\endcsname}%
\let\auto@bib@innerbib\@empty
\bibitem [{\citenamefont {Veldhorst}\ \emph {et~al.}(2014)\citenamefont
  {Veldhorst}, \citenamefont {Hwang}, \citenamefont {Yang}, \citenamefont
  {A.W.}, \citenamefont {de~Ronde}, \citenamefont {Dehollain}, \citenamefont
  {Muhonen}, \citenamefont {Hudson}, \citenamefont {Itoh}, \citenamefont
  {Morello},\ and\ \citenamefont {Dzurak}}]{Veldhorst2014}%
  \BibitemOpen
  \bibfield  {author} {\bibinfo {author} {\bibfnamefont {M.}~\bibnamefont
  {Veldhorst}}, \bibinfo {author} {\bibfnamefont {J.}~\bibnamefont {Hwang}},
  \bibinfo {author} {\bibfnamefont {C.}~\bibnamefont {Yang}}, \bibinfo {author}
  {\bibfnamefont {L.}~\bibnamefont {A.W.}}, \bibinfo {author} {\bibfnamefont
  {B.}~\bibnamefont {de~Ronde}}, \bibinfo {author} {\bibfnamefont {J.~P.}\
  \bibnamefont {Dehollain}}, \bibinfo {author} {\bibfnamefont {J.}~\bibnamefont
  {Muhonen}}, \bibinfo {author} {\bibfnamefont {F.}~\bibnamefont {Hudson}},
  \bibinfo {author} {\bibfnamefont {K.}~\bibnamefont {Itoh}}, \bibinfo {author}
  {\bibfnamefont {A.}~\bibnamefont {Morello}}, \ and\ \bibinfo {author}
  {\bibfnamefont {A.~S.}\ \bibnamefont {Dzurak}},\ }\href@noop {} {\bibfield
  {journal} {\bibinfo  {journal} {Nature}\ }\textbf {\bibinfo {volume} {9}},\
  \bibinfo {pages} {981} (\bibinfo {year} {2014})}\BibitemShut {NoStop}%
\bibitem [{\citenamefont {Kawakami}\ \emph {et~al.}(2014)\citenamefont
  {Kawakami}, \citenamefont {Scarlino}, \citenamefont {Ward}, \citenamefont
  {Braakman}, \citenamefont {Savage}, \citenamefont {Lagally}, \citenamefont
  {Friesen}, \citenamefont {Coppersmith}, \citenamefont {Eriksson},\ and\
  \citenamefont {Vandersypen}}]{kawakami2014electrical}%
  \BibitemOpen
  \bibfield  {author} {\bibinfo {author} {\bibfnamefont {E.}~\bibnamefont
  {Kawakami}}, \bibinfo {author} {\bibfnamefont {P.}~\bibnamefont {Scarlino}},
  \bibinfo {author} {\bibfnamefont {D.}~\bibnamefont {Ward}}, \bibinfo {author}
  {\bibfnamefont {F.}~\bibnamefont {Braakman}}, \bibinfo {author}
  {\bibfnamefont {D.}~\bibnamefont {Savage}}, \bibinfo {author} {\bibfnamefont
  {M.}~\bibnamefont {Lagally}}, \bibinfo {author} {\bibfnamefont
  {M.}~\bibnamefont {Friesen}}, \bibinfo {author} {\bibfnamefont
  {S.}~\bibnamefont {Coppersmith}}, \bibinfo {author} {\bibfnamefont
  {M.}~\bibnamefont {Eriksson}}, \ and\ \bibinfo {author} {\bibfnamefont
  {L.}~\bibnamefont {Vandersypen}},\ }\href@noop {} {\bibfield  {journal}
  {\bibinfo  {journal} {Nature nanotechnology}\ }\textbf {\bibinfo {volume}
  {9}},\ \bibinfo {pages} {666} (\bibinfo {year} {2014})}\BibitemShut {NoStop}%
\bibitem [{\citenamefont {Morello}\ \emph {et~al.}(2010)\citenamefont
  {Morello}, \citenamefont {Pla}, \citenamefont {Zwanenburg}, \citenamefont
  {Chan}, \citenamefont {Tan}, \citenamefont {Huebl}, \citenamefont {Mottonen},
  \citenamefont {Nugroho}, \citenamefont {Yang}, \citenamefont {van Donkelaar},
  \citenamefont {Alves}, \citenamefont {Jamieson}, \citenamefont {Escott},
  \citenamefont {Hollenberg}, \citenamefont {Clark},\ and\ \citenamefont
  {Dzurak}}]{Morello2010}%
  \BibitemOpen
  \bibfield  {author} {\bibinfo {author} {\bibfnamefont {A.}~\bibnamefont
  {Morello}}, \bibinfo {author} {\bibfnamefont {J.~J.}\ \bibnamefont {Pla}},
  \bibinfo {author} {\bibfnamefont {F.~A.}\ \bibnamefont {Zwanenburg}},
  \bibinfo {author} {\bibfnamefont {K.~W.}\ \bibnamefont {Chan}}, \bibinfo
  {author} {\bibfnamefont {K.~Y.}\ \bibnamefont {Tan}}, \bibinfo {author}
  {\bibfnamefont {H.}~\bibnamefont {Huebl}}, \bibinfo {author} {\bibfnamefont
  {M.}~\bibnamefont {Mottonen}}, \bibinfo {author} {\bibfnamefont {C.~D.}\
  \bibnamefont {Nugroho}}, \bibinfo {author} {\bibfnamefont {C.}~\bibnamefont
  {Yang}}, \bibinfo {author} {\bibfnamefont {J.~A.}\ \bibnamefont {van
  Donkelaar}}, \bibinfo {author} {\bibfnamefont {A.~D.~C.}\ \bibnamefont
  {Alves}}, \bibinfo {author} {\bibfnamefont {D.~N.}\ \bibnamefont {Jamieson}},
  \bibinfo {author} {\bibfnamefont {C.~C.}\ \bibnamefont {Escott}}, \bibinfo
  {author} {\bibfnamefont {L.~C.~L.}\ \bibnamefont {Hollenberg}}, \bibinfo
  {author} {\bibfnamefont {R.~G.}\ \bibnamefont {Clark}}, \ and\ \bibinfo
  {author} {\bibfnamefont {A.~S.}\ \bibnamefont {Dzurak}},\ }\href@noop {}
  {\bibfield  {journal} {\bibinfo  {journal} {Nature}\ }\textbf {\bibinfo
  {volume} {467}},\ \bibinfo {pages} {687} (\bibinfo {year}
  {2010})}\BibitemShut {NoStop}%
\bibitem [{\citenamefont {Veldhorst}\ \emph {et~al.}(2015)\citenamefont
  {Veldhorst}, \citenamefont {Yang}, \citenamefont {Hwang}, \citenamefont
  {Huang}, \citenamefont {Dehollain}, \citenamefont {Muhonen}, \citenamefont
  {Simmons}, \citenamefont {Laucht}, \citenamefont {Hudson}, \citenamefont
  {Itoh}, \citenamefont {Morello},\ and\ \citenamefont
  {Dzurak}}]{Veldhorst2015}%
  \BibitemOpen
  \bibfield  {author} {\bibinfo {author} {\bibfnamefont {M.}~\bibnamefont
  {Veldhorst}}, \bibinfo {author} {\bibfnamefont {C.~H.}\ \bibnamefont {Yang}},
  \bibinfo {author} {\bibfnamefont {J.~C.~C.}\ \bibnamefont {Hwang}}, \bibinfo
  {author} {\bibfnamefont {W.}~\bibnamefont {Huang}}, \bibinfo {author}
  {\bibfnamefont {J.~P.}\ \bibnamefont {Dehollain}}, \bibinfo {author}
  {\bibfnamefont {J.~T.}\ \bibnamefont {Muhonen}}, \bibinfo {author}
  {\bibfnamefont {S.}~\bibnamefont {Simmons}}, \bibinfo {author} {\bibfnamefont
  {A.}~\bibnamefont {Laucht}}, \bibinfo {author} {\bibfnamefont {F.~E.}\
  \bibnamefont {Hudson}}, \bibinfo {author} {\bibfnamefont {K.~M.}\
  \bibnamefont {Itoh}}, \bibinfo {author} {\bibfnamefont {A.}~\bibnamefont
  {Morello}}, \ and\ \bibinfo {author} {\bibfnamefont {A.~S.}\ \bibnamefont
  {Dzurak}},\ }\href@noop {} {\bibfield  {journal} {\bibinfo  {journal}
  {Nature}\ }\textbf {\bibinfo {volume} {526}},\ \bibinfo {pages} {410}
  (\bibinfo {year} {2015})}\BibitemShut {NoStop}%
\bibitem [{\citenamefont {Kawakami}\ \emph {et~al.}(2016)\citenamefont
  {Kawakami}, \citenamefont {Jullien}, \citenamefont {Scarlino}, \citenamefont
  {Ward}, \citenamefont {Savage}, \citenamefont {Lagally}, \citenamefont
  {Dobrovitski}, \citenamefont {Friesen}, \citenamefont {Coppersmith},
  \citenamefont {Eriksson},\ and\ \citenamefont {Vandersypen}}]{Kawakami2016}%
  \BibitemOpen
  \bibfield  {author} {\bibinfo {author} {\bibfnamefont {E.}~\bibnamefont
  {Kawakami}}, \bibinfo {author} {\bibfnamefont {T.}~\bibnamefont {Jullien}},
  \bibinfo {author} {\bibfnamefont {P.}~\bibnamefont {Scarlino}}, \bibinfo
  {author} {\bibfnamefont {D.~R.}\ \bibnamefont {Ward}}, \bibinfo {author}
  {\bibfnamefont {D.~E.}\ \bibnamefont {Savage}}, \bibinfo {author}
  {\bibfnamefont {M.~G.}\ \bibnamefont {Lagally}}, \bibinfo {author}
  {\bibfnamefont {V.~V.}\ \bibnamefont {Dobrovitski}}, \bibinfo {author}
  {\bibfnamefont {M.}~\bibnamefont {Friesen}}, \bibinfo {author} {\bibfnamefont
  {S.~N.}\ \bibnamefont {Coppersmith}}, \bibinfo {author} {\bibfnamefont
  {M.~A.}\ \bibnamefont {Eriksson}}, \ and\ \bibinfo {author} {\bibfnamefont
  {L.~M.~K.}\ \bibnamefont {Vandersypen}},\ }\href {\doibase
  10.1073/pnas.1603251113} {\bibfield  {journal} {\bibinfo  {journal}
  {Proceedings of the National Academy of Sciences}\ }\textbf {\bibinfo
  {volume} {113}},\ \bibinfo {pages} {11738} (\bibinfo {year}
  {2016})}\BibitemShut {NoStop}%
\bibitem [{\citenamefont {Zajac}\ \emph {et~al.}(2017)\citenamefont {Zajac},
  \citenamefont {Sigillito}, \citenamefont {Russ}, \citenamefont {Borjans},
  \citenamefont {Taylor}, \citenamefont {Burkard},\ and\ \citenamefont
  {Petta}}]{zajac2017resonantly}%
  \BibitemOpen
  \bibfield  {author} {\bibinfo {author} {\bibfnamefont {D.}~\bibnamefont
  {Zajac}}, \bibinfo {author} {\bibfnamefont {A.}~\bibnamefont {Sigillito}},
  \bibinfo {author} {\bibfnamefont {M.}~\bibnamefont {Russ}}, \bibinfo {author}
  {\bibfnamefont {F.}~\bibnamefont {Borjans}}, \bibinfo {author} {\bibfnamefont
  {J.}~\bibnamefont {Taylor}}, \bibinfo {author} {\bibfnamefont
  {G.}~\bibnamefont {Burkard}}, \ and\ \bibinfo {author} {\bibfnamefont
  {J.}~\bibnamefont {Petta}},\ }\href@noop {} {\bibfield  {journal} {\bibinfo
  {journal} {Science}\ ,\ \bibinfo {pages} {eaao5965}} (\bibinfo {year}
  {2017})}\BibitemShut {NoStop}%
\bibitem [{\citenamefont {Yoneda}\ \emph {et~al.}(2017)\citenamefont {Yoneda},
  \citenamefont {Takeda}, \citenamefont {Otsuka}, \citenamefont {Nakajima},
  \citenamefont {Delbecq}, \citenamefont {Allison}, \citenamefont {Honda},
  \citenamefont {Kodera}, \citenamefont {Oda}, \citenamefont {Hoshi},
  \citenamefont {Usami}, \citenamefont {Itoh},\ and\ \citenamefont
  {Tarucha}}]{yoneda2017quantum}%
  \BibitemOpen
  \bibfield  {author} {\bibinfo {author} {\bibfnamefont {J.}~\bibnamefont
  {Yoneda}}, \bibinfo {author} {\bibfnamefont {K.}~\bibnamefont {Takeda}},
  \bibinfo {author} {\bibfnamefont {T.}~\bibnamefont {Otsuka}}, \bibinfo
  {author} {\bibfnamefont {T.}~\bibnamefont {Nakajima}}, \bibinfo {author}
  {\bibfnamefont {M.~R.}\ \bibnamefont {Delbecq}}, \bibinfo {author}
  {\bibfnamefont {G.}~\bibnamefont {Allison}}, \bibinfo {author} {\bibfnamefont
  {T.}~\bibnamefont {Honda}}, \bibinfo {author} {\bibfnamefont
  {T.}~\bibnamefont {Kodera}}, \bibinfo {author} {\bibfnamefont
  {S.}~\bibnamefont {Oda}}, \bibinfo {author} {\bibfnamefont {Y.}~\bibnamefont
  {Hoshi}}, \bibinfo {author} {\bibfnamefont {N.}~\bibnamefont {Usami}},
  \bibinfo {author} {\bibfnamefont {K.~M.}\ \bibnamefont {Itoh}}, \ and\
  \bibinfo {author} {\bibfnamefont {S.}~\bibnamefont {Tarucha}},\ }\href@noop
  {} {\bibfield  {journal} {\bibinfo  {journal} {Nature nanotechnology}\ ,\
  \bibinfo {pages} {1}} (\bibinfo {year} {2017})}\BibitemShut {NoStop}%
\bibitem [{\citenamefont {Watson}()}]{Watson2017}%
  \BibitemOpen
  \bibfield  {author} {\bibinfo {author} {\bibfnamefont {T.~F.}\ \bibnamefont
  {Watson}},\ }\href@noop {} {\enquote {\bibinfo {title} {A programmable
  two-qubit quantum processor in silicon},}\ }\bibinfo {note}
  {ArXiv:1708.04214}\BibitemShut {NoStop}%
\bibitem [{\citenamefont {DiVincenzo}(2000)}]{DiVincenzo2000}%
  \BibitemOpen
  \bibfield  {author} {\bibinfo {author} {\bibfnamefont {D.~P.}\ \bibnamefont
  {DiVincenzo}},\ }\bibfield  {booktitle} {\emph {\bibinfo {booktitle}
  {Fortschritte Der Physik-Progress of Physics}},\ }\href {<Go to
  ISI>://WOS:000165241900003} {\ \textbf {\bibinfo {volume} {48}},\ \bibinfo
  {pages} {771} (\bibinfo {year} {2000})}\BibitemShut {NoStop}%
\bibitem [{\citenamefont {Vandersypen}\ \emph {et~al.}(2017)\citenamefont
  {Vandersypen}, \citenamefont {Bluhm}, \citenamefont {Clarke}, \citenamefont
  {Dzurak}, \citenamefont {Ishihara}, \citenamefont {Morello}, \citenamefont
  {Reilly}, \citenamefont {Schreiber},\ and\ \citenamefont
  {Veldhorst}}]{Vandersypen2017}%
  \BibitemOpen
  \bibfield  {author} {\bibinfo {author} {\bibfnamefont {L.~M.~K.}\
  \bibnamefont {Vandersypen}}, \bibinfo {author} {\bibfnamefont
  {H.}~\bibnamefont {Bluhm}}, \bibinfo {author} {\bibfnamefont {J.~S.}\
  \bibnamefont {Clarke}}, \bibinfo {author} {\bibfnamefont {A.~S.}\
  \bibnamefont {Dzurak}}, \bibinfo {author} {\bibfnamefont {R.}~\bibnamefont
  {Ishihara}}, \bibinfo {author} {\bibfnamefont {A.}~\bibnamefont {Morello}},
  \bibinfo {author} {\bibfnamefont {D.~J.}\ \bibnamefont {Reilly}}, \bibinfo
  {author} {\bibfnamefont {L.~R.}\ \bibnamefont {Schreiber}}, \ and\ \bibinfo
  {author} {\bibfnamefont {M.}~\bibnamefont {Veldhorst}},\ }\href {\doibase
  10.1038/s41534-017-0038-y} {\bibfield  {journal} {\bibinfo  {journal} {npj
  Quantum Information}\ }\textbf {\bibinfo {volume} {3}},\ \bibinfo {pages}
  {34} (\bibinfo {year} {2017})}\BibitemShut {NoStop}%
\bibitem [{\citenamefont {Veldhorst}\ \emph {et~al.}(2017)\citenamefont
  {Veldhorst}, \citenamefont {Eenink}, \citenamefont {Yang},\ and\
  \citenamefont {Dzurak}}]{Veldhorst2017}%
  \BibitemOpen
  \bibfield  {author} {\bibinfo {author} {\bibfnamefont {M.}~\bibnamefont
  {Veldhorst}}, \bibinfo {author} {\bibfnamefont {H.~G.~J.}\ \bibnamefont
  {Eenink}}, \bibinfo {author} {\bibfnamefont {C.~H.}\ \bibnamefont {Yang}}, \
  and\ \bibinfo {author} {\bibfnamefont {A.~S.}\ \bibnamefont {Dzurak}},\
  }\href {\doibase 10.1038/s41467-017-01905-6} {\bibfield  {journal} {\bibinfo
  {journal} {Nature Communications}\ }\textbf {\bibinfo {volume} {8}},\
  \bibinfo {pages} {1766} (\bibinfo {year} {2017})}\BibitemShut {NoStop}%
\bibitem [{\citenamefont {Li}()}]{Li2017}%
  \BibitemOpen
  \bibfield  {author} {\bibinfo {author} {\bibfnamefont {R.}~\bibnamefont
  {Li}},\ }\href@noop {} {\enquote {\bibinfo {title} {A crossbar network for
  silicon quantum dot qubits},}\ }\bibinfo {note}
  {ArXiv:1711.03807}\BibitemShut {NoStop}%
\bibitem [{\citenamefont {Schoelkopf}\ \emph {et~al.}(1998)\citenamefont
  {Schoelkopf}, \citenamefont {Wahlgren}, \citenamefont {Kozhevnikov},
  \citenamefont {Delsing},\ and\ \citenamefont {Prober}}]{schoelkopf1998radio}%
  \BibitemOpen
  \bibfield  {author} {\bibinfo {author} {\bibfnamefont {R.}~\bibnamefont
  {Schoelkopf}}, \bibinfo {author} {\bibfnamefont {P.}~\bibnamefont
  {Wahlgren}}, \bibinfo {author} {\bibfnamefont {A.}~\bibnamefont
  {Kozhevnikov}}, \bibinfo {author} {\bibfnamefont {P.}~\bibnamefont
  {Delsing}}, \ and\ \bibinfo {author} {\bibfnamefont {D.}~\bibnamefont
  {Prober}},\ }\href@noop {} {\bibfield  {journal} {\bibinfo  {journal}
  {Science}\ }\textbf {\bibinfo {volume} {280}},\ \bibinfo {pages} {1238}
  (\bibinfo {year} {1998})}\BibitemShut {NoStop}%
\bibitem [{\citenamefont {Brenning}\ \emph {et~al.}(2006)\citenamefont
  {Brenning}, \citenamefont {Kafanov}, \citenamefont {Duty}, \citenamefont
  {Kubatkin},\ and\ \citenamefont {Delsing}}]{brenning2006ultrasensitive}%
  \BibitemOpen
  \bibfield  {author} {\bibinfo {author} {\bibfnamefont {H.}~\bibnamefont
  {Brenning}}, \bibinfo {author} {\bibfnamefont {S.}~\bibnamefont {Kafanov}},
  \bibinfo {author} {\bibfnamefont {T.}~\bibnamefont {Duty}}, \bibinfo {author}
  {\bibfnamefont {S.}~\bibnamefont {Kubatkin}}, \ and\ \bibinfo {author}
  {\bibfnamefont {P.}~\bibnamefont {Delsing}},\ }\href@noop {} {\bibfield
  {journal} {\bibinfo  {journal} {Journal of Applied Physics}\ }\textbf
  {\bibinfo {volume} {100}},\ \bibinfo {pages} {114321} (\bibinfo {year}
  {2006})}\BibitemShut {NoStop}%
\bibitem [{\citenamefont {Reilly}\ \emph {et~al.}(2007)\citenamefont {Reilly},
  \citenamefont {Marcus}, \citenamefont {Hanson},\ and\ \citenamefont
  {Gossard}}]{reilly2007fast}%
  \BibitemOpen
  \bibfield  {author} {\bibinfo {author} {\bibfnamefont {D.}~\bibnamefont
  {Reilly}}, \bibinfo {author} {\bibfnamefont {C.}~\bibnamefont {Marcus}},
  \bibinfo {author} {\bibfnamefont {M.}~\bibnamefont {Hanson}}, \ and\ \bibinfo
  {author} {\bibfnamefont {A.}~\bibnamefont {Gossard}},\ }\href@noop {}
  {\bibfield  {journal} {\bibinfo  {journal} {Applied Physics Letters}\
  }\textbf {\bibinfo {volume} {91}},\ \bibinfo {pages} {162101} (\bibinfo
  {year} {2007})}\BibitemShut {NoStop}%
\bibitem [{\citenamefont {Wallraff}\ \emph {et~al.}(2004)\citenamefont
  {Wallraff}, \citenamefont {Schuster}, \citenamefont {Blais}, \citenamefont
  {Frunzio}, \citenamefont {Huang}, \citenamefont {Kumar}, \citenamefont
  {Girvin},\ and\ \citenamefont {Schoelkopf}}]{Wallraff2004}%
  \BibitemOpen
  \bibfield  {author} {\bibinfo {author} {\bibfnamefont {A.}~\bibnamefont
  {Wallraff}}, \bibinfo {author} {\bibfnamefont {D.~I.}\ \bibnamefont
  {Schuster}}, \bibinfo {author} {\bibfnamefont {A.}~\bibnamefont {Blais}},
  \bibinfo {author} {\bibfnamefont {L.}~\bibnamefont {Frunzio}}, \bibinfo
  {author} {\bibfnamefont {J.}~\bibnamefont {Huang}, \bibfnamefont
  {R.~S.and~Majer}}, \bibinfo {author} {\bibfnamefont {S.}~\bibnamefont
  {Kumar}}, \bibinfo {author} {\bibfnamefont {S.~M.}\ \bibnamefont {Girvin}}, \
  and\ \bibinfo {author} {\bibfnamefont {R.~J.}\ \bibnamefont {Schoelkopf}},\
  }\href@noop {} {\bibfield  {journal} {\bibinfo  {journal} {Nature}\ }\textbf
  {\bibinfo {volume} {431}},\ \bibinfo {pages} {162} (\bibinfo {year}
  {2004})}\BibitemShut {NoStop}%
\bibitem [{\citenamefont {Petersson}\ \emph {et~al.}(2012)\citenamefont
  {Petersson}, \citenamefont {McFaul}, \citenamefont {Schroer}, \citenamefont
  {Jung}, \citenamefont {Taylor}, \citenamefont {Houck},\ and\ \citenamefont
  {Petta}}]{Petersson2012}%
  \BibitemOpen
  \bibfield  {author} {\bibinfo {author} {\bibfnamefont {K.~D.}\ \bibnamefont
  {Petersson}}, \bibinfo {author} {\bibfnamefont {L.~W.}\ \bibnamefont
  {McFaul}}, \bibinfo {author} {\bibfnamefont {M.~D.}\ \bibnamefont {Schroer}},
  \bibinfo {author} {\bibfnamefont {M.}~\bibnamefont {Jung}}, \bibinfo {author}
  {\bibfnamefont {J.~M.}\ \bibnamefont {Taylor}}, \bibinfo {author}
  {\bibfnamefont {A.~A.}\ \bibnamefont {Houck}}, \ and\ \bibinfo {author}
  {\bibfnamefont {J.~R.}\ \bibnamefont {Petta}},\ }\href@noop {} {\bibfield
  {journal} {\bibinfo  {journal} {Nature}\ }\textbf {\bibinfo {volume} {490}},\
  \bibinfo {pages} {380} (\bibinfo {year} {2012})}\BibitemShut {NoStop}%
\bibitem [{\citenamefont {Stockklauser}\ \emph {et~al.}(2017)\citenamefont
  {Stockklauser}, \citenamefont {Scarlino}, \citenamefont {Koski},
  \citenamefont {Gasparinetti}, \citenamefont {Andersen}, \citenamefont
  {Reichl}, \citenamefont {Wegscheider}, \citenamefont {Ihn}, \citenamefont
  {Ensslin},\ and\ \citenamefont {Wallraff}}]{Stockklauser2017}%
  \BibitemOpen
  \bibfield  {author} {\bibinfo {author} {\bibfnamefont {A.}~\bibnamefont
  {Stockklauser}}, \bibinfo {author} {\bibfnamefont {P.}~\bibnamefont
  {Scarlino}}, \bibinfo {author} {\bibfnamefont {J.~V.}\ \bibnamefont {Koski}},
  \bibinfo {author} {\bibfnamefont {S.}~\bibnamefont {Gasparinetti}}, \bibinfo
  {author} {\bibfnamefont {C.~K.}\ \bibnamefont {Andersen}}, \bibinfo {author}
  {\bibfnamefont {C.}~\bibnamefont {Reichl}}, \bibinfo {author} {\bibfnamefont
  {W.}~\bibnamefont {Wegscheider}}, \bibinfo {author} {\bibfnamefont
  {T.}~\bibnamefont {Ihn}}, \bibinfo {author} {\bibfnamefont {K.}~\bibnamefont
  {Ensslin}}, \ and\ \bibinfo {author} {\bibfnamefont {A.}~\bibnamefont
  {Wallraff}},\ }\href {\doibase 10.1103/PhysRevX.7.011030} {\bibfield
  {journal} {\bibinfo  {journal} {Phys. Rev. X}\ }\textbf {\bibinfo {volume}
  {7}},\ \bibinfo {pages} {011030} (\bibinfo {year} {2017})}\BibitemShut
  {NoStop}%
\bibitem [{\citenamefont {Mi}\ \emph {et~al.}(2016)\citenamefont {Mi},
  \citenamefont {Cady}, \citenamefont {Zajac}, \citenamefont {Deelman},\ and\
  \citenamefont {Petta}}]{mi2016strong}%
  \BibitemOpen
  \bibfield  {author} {\bibinfo {author} {\bibfnamefont {X.}~\bibnamefont
  {Mi}}, \bibinfo {author} {\bibfnamefont {J.}~\bibnamefont {Cady}}, \bibinfo
  {author} {\bibfnamefont {D.}~\bibnamefont {Zajac}}, \bibinfo {author}
  {\bibfnamefont {P.}~\bibnamefont {Deelman}}, \ and\ \bibinfo {author}
  {\bibfnamefont {J.}~\bibnamefont {Petta}},\ }\href@noop {} {\bibfield
  {journal} {\bibinfo  {journal} {Science}\ ,\ \bibinfo {pages} {aal2469}}
  (\bibinfo {year} {2016})}\BibitemShut {NoStop}%
\bibitem [{\citenamefont {Colless}\ \emph {et~al.}(2013)\citenamefont
  {Colless}, \citenamefont {Mahoney}, \citenamefont {Hornibrook}, \citenamefont
  {Doherty}, \citenamefont {Lu}, \citenamefont {Gossard},\ and\ \citenamefont
  {Reilly}}]{colless2013dispersive}%
  \BibitemOpen
  \bibfield  {author} {\bibinfo {author} {\bibfnamefont {J.}~\bibnamefont
  {Colless}}, \bibinfo {author} {\bibfnamefont {A.}~\bibnamefont {Mahoney}},
  \bibinfo {author} {\bibfnamefont {J.}~\bibnamefont {Hornibrook}}, \bibinfo
  {author} {\bibfnamefont {A.}~\bibnamefont {Doherty}}, \bibinfo {author}
  {\bibfnamefont {H.}~\bibnamefont {Lu}}, \bibinfo {author} {\bibfnamefont
  {A.}~\bibnamefont {Gossard}}, \ and\ \bibinfo {author} {\bibfnamefont
  {D.}~\bibnamefont {Reilly}},\ }\href@noop {} {\bibfield  {journal} {\bibinfo
  {journal} {Physical review letters}\ }\textbf {\bibinfo {volume} {110}},\
  \bibinfo {pages} {046805} (\bibinfo {year} {2013})}\BibitemShut {NoStop}%
\bibitem [{\citenamefont {Gonzalez-Zalba}\ \emph {et~al.}(2015)\citenamefont
  {Gonzalez-Zalba}, \citenamefont {Barraud}, \citenamefont {Ferguson},\ and\
  \citenamefont {Betz}}]{gonzalez2015probing}%
  \BibitemOpen
  \bibfield  {author} {\bibinfo {author} {\bibfnamefont {M.}~\bibnamefont
  {Gonzalez-Zalba}}, \bibinfo {author} {\bibfnamefont {S.}~\bibnamefont
  {Barraud}}, \bibinfo {author} {\bibfnamefont {A.}~\bibnamefont {Ferguson}}, \
  and\ \bibinfo {author} {\bibfnamefont {A.}~\bibnamefont {Betz}},\ }\href@noop
  {} {\bibfield  {journal} {\bibinfo  {journal} {Nature communications}\
  }\textbf {\bibinfo {volume} {6}},\ \bibinfo {pages} {6084} (\bibinfo {year}
  {2015})}\BibitemShut {NoStop}%
\bibitem [{\citenamefont {House}\ \emph {et~al.}(2015)\citenamefont {House},
  \citenamefont {Kobayashi}, \citenamefont {Weber}, \citenamefont {Hile},
  \citenamefont {Watson}, \citenamefont {Van Der~Heijden}, \citenamefont
  {Rogge},\ and\ \citenamefont {Simmons}}]{house2015radio}%
  \BibitemOpen
  \bibfield  {author} {\bibinfo {author} {\bibfnamefont {M.}~\bibnamefont
  {House}}, \bibinfo {author} {\bibfnamefont {T.}~\bibnamefont {Kobayashi}},
  \bibinfo {author} {\bibfnamefont {B.}~\bibnamefont {Weber}}, \bibinfo
  {author} {\bibfnamefont {S.}~\bibnamefont {Hile}}, \bibinfo {author}
  {\bibfnamefont {T.}~\bibnamefont {Watson}}, \bibinfo {author} {\bibfnamefont
  {J.}~\bibnamefont {Van Der~Heijden}}, \bibinfo {author} {\bibfnamefont
  {S.}~\bibnamefont {Rogge}}, \ and\ \bibinfo {author} {\bibfnamefont
  {M.}~\bibnamefont {Simmons}},\ }\href@noop {} {\bibfield  {journal} {\bibinfo
   {journal} {Nature communications}\ }\textbf {\bibinfo {volume} {6}},\
  \bibinfo {pages} {8848} (\bibinfo {year} {2015})}\BibitemShut {NoStop}%
\bibitem [{\citenamefont {Ares}\ \emph {et~al.}(2016)\citenamefont {Ares},
  \citenamefont {Schupp}, \citenamefont {Mavalankar}, \citenamefont {Rogers},
  \citenamefont {Griffiths}, \citenamefont {Jones}, \citenamefont {Farrer},
  \citenamefont {Ritchie}, \citenamefont {Smith}, \citenamefont {Cottet},
  \citenamefont {Briggs},\ and\ \citenamefont {Laird}}]{Ares2016}%
  \BibitemOpen
  \bibfield  {author} {\bibinfo {author} {\bibfnamefont {N.}~\bibnamefont
  {Ares}}, \bibinfo {author} {\bibfnamefont {F.~J.}\ \bibnamefont {Schupp}},
  \bibinfo {author} {\bibfnamefont {A.}~\bibnamefont {Mavalankar}}, \bibinfo
  {author} {\bibfnamefont {G.}~\bibnamefont {Rogers}}, \bibinfo {author}
  {\bibfnamefont {J.}~\bibnamefont {Griffiths}}, \bibinfo {author}
  {\bibfnamefont {G.~A.~C.}\ \bibnamefont {Jones}}, \bibinfo {author}
  {\bibfnamefont {I.}~\bibnamefont {Farrer}}, \bibinfo {author} {\bibfnamefont
  {D.~A.}\ \bibnamefont {Ritchie}}, \bibinfo {author} {\bibfnamefont {C.~G.}\
  \bibnamefont {Smith}}, \bibinfo {author} {\bibfnamefont {A.}~\bibnamefont
  {Cottet}}, \bibinfo {author} {\bibfnamefont {G.~A.~D.}\ \bibnamefont
  {Briggs}}, \ and\ \bibinfo {author} {\bibfnamefont {E.~A.}\ \bibnamefont
  {Laird}},\ }\href {\doibase 10.1103/PhysRevApplied.5.034011} {\bibfield
  {journal} {\bibinfo  {journal} {Phys. Rev. Applied}\ }\textbf {\bibinfo
  {volume} {5}},\ \bibinfo {pages} {034011} (\bibinfo {year}
  {2016})}\BibitemShut {NoStop}%
\bibitem [{\citenamefont {Crippa}\ \emph {et~al.}(2017)\citenamefont {Crippa},
  \citenamefont {Maurand}, \citenamefont {Kotekar-Patil}, \citenamefont
  {Corna}, \citenamefont {Bohuslavskyi}, \citenamefont {Orlov}, \citenamefont
  {Fay}, \citenamefont {Lavi{\'e}ville}, \citenamefont {Barraud}, \citenamefont
  {Vinet}, \citenamefont {Sanquer}, \citenamefont {Franceschi},\ and\
  \citenamefont {Jehl}}]{crippa2017level}%
  \BibitemOpen
  \bibfield  {author} {\bibinfo {author} {\bibfnamefont {A.}~\bibnamefont
  {Crippa}}, \bibinfo {author} {\bibfnamefont {R.}~\bibnamefont {Maurand}},
  \bibinfo {author} {\bibfnamefont {D.}~\bibnamefont {Kotekar-Patil}}, \bibinfo
  {author} {\bibfnamefont {A.}~\bibnamefont {Corna}}, \bibinfo {author}
  {\bibfnamefont {H.}~\bibnamefont {Bohuslavskyi}}, \bibinfo {author}
  {\bibfnamefont {A.~O.}\ \bibnamefont {Orlov}}, \bibinfo {author}
  {\bibfnamefont {P.}~\bibnamefont {Fay}}, \bibinfo {author} {\bibfnamefont
  {R.}~\bibnamefont {Lavi{\'e}ville}}, \bibinfo {author} {\bibfnamefont
  {S.}~\bibnamefont {Barraud}}, \bibinfo {author} {\bibfnamefont
  {M.}~\bibnamefont {Vinet}}, \bibinfo {author} {\bibfnamefont
  {M.}~\bibnamefont {Sanquer}}, \bibinfo {author} {\bibfnamefont {S.~D.}\
  \bibnamefont {Franceschi}}, \ and\ \bibinfo {author} {\bibfnamefont
  {X.}~\bibnamefont {Jehl}},\ }\href@noop {} {\bibfield  {journal} {\bibinfo
  {journal} {Nano letters}\ }\textbf {\bibinfo {volume} {17}},\ \bibinfo
  {pages} {1001} (\bibinfo {year} {2017})}\BibitemShut {NoStop}%
\bibitem [{\citenamefont {Ashoori}\ \emph {et~al.}(1992)\citenamefont
  {Ashoori}, \citenamefont {Stormer}, \citenamefont {Weiner}, \citenamefont
  {Pfeiffer}, \citenamefont {Pearton}, \citenamefont {Baldwin},\ and\
  \citenamefont {West}}]{Ashoori1992}%
  \BibitemOpen
  \bibfield  {author} {\bibinfo {author} {\bibfnamefont {R.~C.}\ \bibnamefont
  {Ashoori}}, \bibinfo {author} {\bibfnamefont {H.~L.}\ \bibnamefont
  {Stormer}}, \bibinfo {author} {\bibfnamefont {J.~S.}\ \bibnamefont {Weiner}},
  \bibinfo {author} {\bibfnamefont {L.~N.}\ \bibnamefont {Pfeiffer}}, \bibinfo
  {author} {\bibfnamefont {S.~J.}\ \bibnamefont {Pearton}}, \bibinfo {author}
  {\bibfnamefont {K.~W.}\ \bibnamefont {Baldwin}}, \ and\ \bibinfo {author}
  {\bibfnamefont {K.~W.}\ \bibnamefont {West}},\ }\href@noop {} {\bibfield
  {journal} {\bibinfo  {journal} {Phys. Rev. Lett.}\ }\textbf {\bibinfo
  {volume} {68}},\ \bibinfo {pages} {3088} (\bibinfo {year}
  {1992})}\BibitemShut {NoStop}%
\bibitem [{\citenamefont {Ciccarelli}\ and\ \citenamefont
  {Ferguson}(2011)}]{ciccarelli2011impedance}%
  \BibitemOpen
  \bibfield  {author} {\bibinfo {author} {\bibfnamefont {C.}~\bibnamefont
  {Ciccarelli}}\ and\ \bibinfo {author} {\bibfnamefont {A.}~\bibnamefont
  {Ferguson}},\ }\href@noop {} {\bibfield  {journal} {\bibinfo  {journal} {New
  Journal of Physics}\ }\textbf {\bibinfo {volume} {13}},\ \bibinfo {pages}
  {093015} (\bibinfo {year} {2011})}\BibitemShut {NoStop}%
\bibitem [{\citenamefont {Persson}\ \emph {et~al.}(2010)\citenamefont
  {Persson}, \citenamefont {Wilson}, \citenamefont {Sandberg}, \citenamefont
  {Johansson},\ and\ \citenamefont {Delsing}}]{persson2010excess}%
  \BibitemOpen
  \bibfield  {author} {\bibinfo {author} {\bibfnamefont {F.}~\bibnamefont
  {Persson}}, \bibinfo {author} {\bibfnamefont {C.}~\bibnamefont {Wilson}},
  \bibinfo {author} {\bibfnamefont {M.}~\bibnamefont {Sandberg}}, \bibinfo
  {author} {\bibfnamefont {G.}~\bibnamefont {Johansson}}, \ and\ \bibinfo
  {author} {\bibfnamefont {P.}~\bibnamefont {Delsing}},\ }\href@noop {}
  {\bibfield  {journal} {\bibinfo  {journal} {Nano letters}\ }\textbf {\bibinfo
  {volume} {10}},\ \bibinfo {pages} {953} (\bibinfo {year} {2010})}\BibitemShut
  {NoStop}%
\bibitem [{\citenamefont {Mizuta}\ \emph {et~al.}(2017)\citenamefont {Mizuta},
  \citenamefont {Otxoa}, \citenamefont {Betz},\ and\ \citenamefont
  {Gonzalez-Zalba}}]{Mizuta2017}%
  \BibitemOpen
  \bibfield  {author} {\bibinfo {author} {\bibfnamefont {R.}~\bibnamefont
  {Mizuta}}, \bibinfo {author} {\bibfnamefont {R.~M.}\ \bibnamefont {Otxoa}},
  \bibinfo {author} {\bibfnamefont {A.~C.}\ \bibnamefont {Betz}}, \ and\
  \bibinfo {author} {\bibfnamefont {M.~F.}\ \bibnamefont {Gonzalez-Zalba}},\
  }\href {\doibase 10.1103/PhysRevB.95.045414} {\bibfield  {journal} {\bibinfo
  {journal} {Phys. Rev. B}\ }\textbf {\bibinfo {volume} {95}},\ \bibinfo
  {pages} {045414} (\bibinfo {year} {2017})}\BibitemShut {NoStop}%
\bibitem [{\citenamefont {Betz}\ \emph {et~al.}(2014)\citenamefont {Betz},
  \citenamefont {Barraud}, \citenamefont {Wilmart}, \citenamefont {Placais},
  \citenamefont {Jehl}, \citenamefont {Sanquer},\ and\ \citenamefont
  {Gonzalez-Zalba}}]{betz2014high}%
  \BibitemOpen
  \bibfield  {author} {\bibinfo {author} {\bibfnamefont {A.}~\bibnamefont
  {Betz}}, \bibinfo {author} {\bibfnamefont {S.}~\bibnamefont {Barraud}},
  \bibinfo {author} {\bibfnamefont {Q.}~\bibnamefont {Wilmart}}, \bibinfo
  {author} {\bibfnamefont {B.}~\bibnamefont {Placais}}, \bibinfo {author}
  {\bibfnamefont {X.}~\bibnamefont {Jehl}}, \bibinfo {author} {\bibfnamefont
  {M.}~\bibnamefont {Sanquer}}, \ and\ \bibinfo {author} {\bibfnamefont
  {M.}~\bibnamefont {Gonzalez-Zalba}},\ }\href@noop {} {\bibfield  {journal}
  {\bibinfo  {journal} {Applied Physics Letters}\ }\textbf {\bibinfo {volume}
  {104}},\ \bibinfo {pages} {043106} (\bibinfo {year} {2014})}\BibitemShut
  {NoStop}%
\bibitem [{\citenamefont {Voisin}\ \emph {et~al.}(2014)\citenamefont {Voisin},
  \citenamefont {Nguyen}, \citenamefont {Renard}, \citenamefont {Jehl},
  \citenamefont {Barraud}, \citenamefont {Triozon}, \citenamefont {Vinet},
  \citenamefont {Duchemin}, \citenamefont {Niquet}, \citenamefont
  {de~Franceschi} \emph {et~al.}}]{voisin2014few}%
  \BibitemOpen
  \bibfield  {author} {\bibinfo {author} {\bibfnamefont {B.}~\bibnamefont
  {Voisin}}, \bibinfo {author} {\bibfnamefont {V.-H.}\ \bibnamefont {Nguyen}},
  \bibinfo {author} {\bibfnamefont {J.}~\bibnamefont {Renard}}, \bibinfo
  {author} {\bibfnamefont {X.}~\bibnamefont {Jehl}}, \bibinfo {author}
  {\bibfnamefont {S.}~\bibnamefont {Barraud}}, \bibinfo {author} {\bibfnamefont
  {F.}~\bibnamefont {Triozon}}, \bibinfo {author} {\bibfnamefont
  {M.}~\bibnamefont {Vinet}}, \bibinfo {author} {\bibfnamefont
  {I.}~\bibnamefont {Duchemin}}, \bibinfo {author} {\bibfnamefont {Y.-M.}\
  \bibnamefont {Niquet}}, \bibinfo {author} {\bibfnamefont {S.}~\bibnamefont
  {de~Franceschi}},  \emph {et~al.},\ }\href@noop {} {\bibfield  {journal}
  {\bibinfo  {journal} {Nano letters}\ }\textbf {\bibinfo {volume} {14}},\
  \bibinfo {pages} {2094} (\bibinfo {year} {2014})}\BibitemShut {NoStop}%
\bibitem [{\citenamefont {Sellier}\ \emph {et~al.}(2007)\citenamefont
  {Sellier}, \citenamefont {Lansbergen}, \citenamefont {Caro}, \citenamefont
  {Rogge}, \citenamefont {Collaert}, \citenamefont {Ferain}, \citenamefont
  {Jurczak},\ and\ \citenamefont {Biesemans}}]{sellier2007subthreshold}%
  \BibitemOpen
  \bibfield  {author} {\bibinfo {author} {\bibfnamefont {H.}~\bibnamefont
  {Sellier}}, \bibinfo {author} {\bibfnamefont {G.}~\bibnamefont {Lansbergen}},
  \bibinfo {author} {\bibfnamefont {J.}~\bibnamefont {Caro}}, \bibinfo {author}
  {\bibfnamefont {S.}~\bibnamefont {Rogge}}, \bibinfo {author} {\bibfnamefont
  {N.}~\bibnamefont {Collaert}}, \bibinfo {author} {\bibfnamefont
  {I.}~\bibnamefont {Ferain}}, \bibinfo {author} {\bibfnamefont
  {M.}~\bibnamefont {Jurczak}}, \ and\ \bibinfo {author} {\bibfnamefont
  {S.}~\bibnamefont {Biesemans}},\ }\href@noop {} {\bibfield  {journal}
  {\bibinfo  {journal} {Applied physics letters}\ }\textbf {\bibinfo {volume}
  {90}},\ \bibinfo {pages} {073502} (\bibinfo {year} {2007})}\BibitemShut
  {NoStop}%
\bibitem [{\citenamefont {Hornibrook}\ \emph {et~al.}(2014)\citenamefont
  {Hornibrook}, \citenamefont {Colless}, \citenamefont {Mahoney}, \citenamefont
  {Croot}, \citenamefont {Blanvillain}, \citenamefont {Lu}, \citenamefont
  {Gossard},\ and\ \citenamefont {Reilly}}]{hornibrook2014frequency}%
  \BibitemOpen
  \bibfield  {author} {\bibinfo {author} {\bibfnamefont {J.}~\bibnamefont
  {Hornibrook}}, \bibinfo {author} {\bibfnamefont {J.}~\bibnamefont {Colless}},
  \bibinfo {author} {\bibfnamefont {A.}~\bibnamefont {Mahoney}}, \bibinfo
  {author} {\bibfnamefont {X.}~\bibnamefont {Croot}}, \bibinfo {author}
  {\bibfnamefont {S.}~\bibnamefont {Blanvillain}}, \bibinfo {author}
  {\bibfnamefont {H.}~\bibnamefont {Lu}}, \bibinfo {author} {\bibfnamefont
  {A.}~\bibnamefont {Gossard}}, \ and\ \bibinfo {author} {\bibfnamefont
  {D.}~\bibnamefont {Reilly}},\ }\href@noop {} {\bibfield  {journal} {\bibinfo
  {journal} {Applied Physics Letters}\ }\textbf {\bibinfo {volume} {104}},\
  \bibinfo {pages} {103108} (\bibinfo {year} {2014})}\BibitemShut {NoStop}%
\bibitem [{\citenamefont {Stevenson}\ \emph {et~al.}(2002)\citenamefont
  {Stevenson}, \citenamefont {Pellerano}, \citenamefont {Stahle}, \citenamefont
  {Aidala},\ and\ \citenamefont {Schoelkopf}}]{stevenson2002multiplexing}%
  \BibitemOpen
  \bibfield  {author} {\bibinfo {author} {\bibfnamefont {T.~R.}\ \bibnamefont
  {Stevenson}}, \bibinfo {author} {\bibfnamefont {F.}~\bibnamefont
  {Pellerano}}, \bibinfo {author} {\bibfnamefont {C.}~\bibnamefont {Stahle}},
  \bibinfo {author} {\bibfnamefont {K.}~\bibnamefont {Aidala}}, \ and\ \bibinfo
  {author} {\bibfnamefont {R.}~\bibnamefont {Schoelkopf}},\ }\href@noop {}
  {\bibfield  {journal} {\bibinfo  {journal} {Applied physics letters}\
  }\textbf {\bibinfo {volume} {80}},\ \bibinfo {pages} {3012} (\bibinfo {year}
  {2002})}\BibitemShut {NoStop}%
\bibitem [{\citenamefont {Lambert}\ \emph {et~al.}(2014)\citenamefont
  {Lambert}, \citenamefont {Edwards}, \citenamefont {Ciccarelli},\ and\
  \citenamefont {Ferguson}}]{Lambert2014}%
  \BibitemOpen
  \bibfield  {author} {\bibinfo {author} {\bibfnamefont {N.~J.}\ \bibnamefont
  {Lambert}}, \bibinfo {author} {\bibfnamefont {M.}~\bibnamefont {Edwards}},
  \bibinfo {author} {\bibfnamefont {C.}~\bibnamefont {Ciccarelli}}, \ and\
  \bibinfo {author} {\bibfnamefont {A.~J.}\ \bibnamefont {Ferguson}},\ }\href
  {\doibase 10.1021/nl403659x} {\bibfield  {journal} {\bibinfo  {journal} {Nano
  Letters}\ }\textbf {\bibinfo {volume} {14}},\ \bibinfo {pages} {1148}
  (\bibinfo {year} {2014})}\BibitemShut {NoStop}%
\bibitem [{\citenamefont {Chorley}\ \emph {et~al.}(2012)\citenamefont
  {Chorley}, \citenamefont {Wabnig}, \citenamefont {Penfold-Fitch},
  \citenamefont {Petersson}, \citenamefont {Frake}, \citenamefont {Smith},\
  and\ \citenamefont {Buitelaar}}]{chorley2012measuring}%
  \BibitemOpen
  \bibfield  {author} {\bibinfo {author} {\bibfnamefont {S.}~\bibnamefont
  {Chorley}}, \bibinfo {author} {\bibfnamefont {J.}~\bibnamefont {Wabnig}},
  \bibinfo {author} {\bibfnamefont {Z.}~\bibnamefont {Penfold-Fitch}}, \bibinfo
  {author} {\bibfnamefont {K.}~\bibnamefont {Petersson}}, \bibinfo {author}
  {\bibfnamefont {J.}~\bibnamefont {Frake}}, \bibinfo {author} {\bibfnamefont
  {C.}~\bibnamefont {Smith}}, \ and\ \bibinfo {author} {\bibfnamefont
  {M.}~\bibnamefont {Buitelaar}},\ }\href@noop {} {\bibfield  {journal}
  {\bibinfo  {journal} {Physical review letters}\ }\textbf {\bibinfo {volume}
  {108}},\ \bibinfo {pages} {036802} (\bibinfo {year} {2012})}\BibitemShut
  {NoStop}%
\bibitem [{\citenamefont {Petersson}\ \emph {et~al.}(2010)\citenamefont
  {Petersson}, \citenamefont {Smith}, \citenamefont {Anderson}, \citenamefont
  {Atkinson}, \citenamefont {Jones},\ and\ \citenamefont
  {Ritchie}}]{petersson2010charge}%
  \BibitemOpen
  \bibfield  {author} {\bibinfo {author} {\bibfnamefont {K.}~\bibnamefont
  {Petersson}}, \bibinfo {author} {\bibfnamefont {C.}~\bibnamefont {Smith}},
  \bibinfo {author} {\bibfnamefont {D.}~\bibnamefont {Anderson}}, \bibinfo
  {author} {\bibfnamefont {P.}~\bibnamefont {Atkinson}}, \bibinfo {author}
  {\bibfnamefont {G.}~\bibnamefont {Jones}}, \ and\ \bibinfo {author}
  {\bibfnamefont {D.}~\bibnamefont {Ritchie}},\ }\href@noop {} {\bibfield
  {journal} {\bibinfo  {journal} {Nano letters}\ }\textbf {\bibinfo {volume}
  {10}},\ \bibinfo {pages} {2789} (\bibinfo {year} {2010})}\BibitemShut
  {NoStop}%
\bibitem [{\citenamefont {Gonzalez-Zalba}\ \emph {et~al.}(2016)\citenamefont
  {Gonzalez-Zalba}, \citenamefont {Shevchenko}, \citenamefont {Barraud},
  \citenamefont {Johansson}, \citenamefont {Ferguson}, \citenamefont {Nori},\
  and\ \citenamefont {Betz}}]{gonzalez2016gate}%
  \BibitemOpen
  \bibfield  {author} {\bibinfo {author} {\bibfnamefont {M.~F.}\ \bibnamefont
  {Gonzalez-Zalba}}, \bibinfo {author} {\bibfnamefont {S.~N.}\ \bibnamefont
  {Shevchenko}}, \bibinfo {author} {\bibfnamefont {S.}~\bibnamefont {Barraud}},
  \bibinfo {author} {\bibfnamefont {J.~R.}\ \bibnamefont {Johansson}}, \bibinfo
  {author} {\bibfnamefont {A.~J.}\ \bibnamefont {Ferguson}}, \bibinfo {author}
  {\bibfnamefont {F.}~\bibnamefont {Nori}}, \ and\ \bibinfo {author}
  {\bibfnamefont {A.~C.}\ \bibnamefont {Betz}},\ }\href@noop {} {\bibfield
  {journal} {\bibinfo  {journal} {Nano letters}\ }\textbf {\bibinfo {volume}
  {16}},\ \bibinfo {pages} {1614} (\bibinfo {year} {2016})}\BibitemShut
  {NoStop}%
\bibitem [{\citenamefont {Urdampilleta}\ \emph {et~al.}(2015)\citenamefont
  {Urdampilleta}, \citenamefont {Chatterjee}, \citenamefont {Lo}, \citenamefont
  {Kobayashi}, \citenamefont {Mansir}, \citenamefont {Barraud}, \citenamefont
  {Betz}, \citenamefont {Rogge}, \citenamefont {Gonzalez-Zalba},\ and\
  \citenamefont {Morton}}]{urdampilleta2015charge}%
  \BibitemOpen
  \bibfield  {author} {\bibinfo {author} {\bibfnamefont {M.}~\bibnamefont
  {Urdampilleta}}, \bibinfo {author} {\bibfnamefont {A.}~\bibnamefont
  {Chatterjee}}, \bibinfo {author} {\bibfnamefont {C.~C.}\ \bibnamefont {Lo}},
  \bibinfo {author} {\bibfnamefont {T.}~\bibnamefont {Kobayashi}}, \bibinfo
  {author} {\bibfnamefont {J.}~\bibnamefont {Mansir}}, \bibinfo {author}
  {\bibfnamefont {S.}~\bibnamefont {Barraud}}, \bibinfo {author} {\bibfnamefont
  {A.~C.}\ \bibnamefont {Betz}}, \bibinfo {author} {\bibfnamefont
  {S.}~\bibnamefont {Rogge}}, \bibinfo {author} {\bibfnamefont {M.~F.}\
  \bibnamefont {Gonzalez-Zalba}}, \ and\ \bibinfo {author} {\bibfnamefont
  {J.~J.}\ \bibnamefont {Morton}},\ }\href@noop {} {\bibfield  {journal}
  {\bibinfo  {journal} {Physical Review X}\ }\textbf {\bibinfo {volume} {5}},\
  \bibinfo {pages} {031024} (\bibinfo {year} {2015})}\BibitemShut {NoStop}%
\bibitem [{\citenamefont {Cottet}\ \emph {et~al.}(2011)\citenamefont {Cottet},
  \citenamefont {Mora},\ and\ \citenamefont {Kontos}}]{cottet2011mesoscopic}%
  \BibitemOpen
  \bibfield  {author} {\bibinfo {author} {\bibfnamefont {A.}~\bibnamefont
  {Cottet}}, \bibinfo {author} {\bibfnamefont {C.}~\bibnamefont {Mora}}, \ and\
  \bibinfo {author} {\bibfnamefont {T.}~\bibnamefont {Kontos}},\ }\href@noop {}
  {\bibfield  {journal} {\bibinfo  {journal} {Physical Review B}\ }\textbf
  {\bibinfo {volume} {83}},\ \bibinfo {pages} {121311} (\bibinfo {year}
  {2011})}\BibitemShut {NoStop}%
\bibitem [{\citenamefont {Hanson}\ \emph {et~al.}(2007)\citenamefont {Hanson},
  \citenamefont {Kouwenhoven}, \citenamefont {Petta}, \citenamefont {Tarucha},\
  and\ \citenamefont {Vandersypen}}]{hanson2007spins}%
  \BibitemOpen
  \bibfield  {author} {\bibinfo {author} {\bibfnamefont {R.}~\bibnamefont
  {Hanson}}, \bibinfo {author} {\bibfnamefont {L.~P.}\ \bibnamefont
  {Kouwenhoven}}, \bibinfo {author} {\bibfnamefont {J.~R.}\ \bibnamefont
  {Petta}}, \bibinfo {author} {\bibfnamefont {S.}~\bibnamefont {Tarucha}}, \
  and\ \bibinfo {author} {\bibfnamefont {L.~M.}\ \bibnamefont {Vandersypen}},\
  }\href@noop {} {\bibfield  {journal} {\bibinfo  {journal} {Reviews of Modern
  Physics}\ }\textbf {\bibinfo {volume} {79}},\ \bibinfo {pages} {1217}
  (\bibinfo {year} {2007})}\BibitemShut {NoStop}%
\bibitem [{\citenamefont {Roschier}\ \emph {et~al.}(2004)\citenamefont
  {Roschier}, \citenamefont {Hakonen}, \citenamefont {Bladh}, \citenamefont
  {Delsing}, \citenamefont {Lehnert}, \citenamefont {Spietz},\ and\
  \citenamefont {Schoelkopf}}]{roschier2004noise}%
  \BibitemOpen
  \bibfield  {author} {\bibinfo {author} {\bibfnamefont {L.}~\bibnamefont
  {Roschier}}, \bibinfo {author} {\bibfnamefont {P.}~\bibnamefont {Hakonen}},
  \bibinfo {author} {\bibfnamefont {K.}~\bibnamefont {Bladh}}, \bibinfo
  {author} {\bibfnamefont {P.}~\bibnamefont {Delsing}}, \bibinfo {author}
  {\bibfnamefont {K.}~\bibnamefont {Lehnert}}, \bibinfo {author} {\bibfnamefont
  {L.}~\bibnamefont {Spietz}}, \ and\ \bibinfo {author} {\bibfnamefont
  {R.}~\bibnamefont {Schoelkopf}},\ }\href@noop {} {\bibfield  {journal}
  {\bibinfo  {journal} {Journal of applied physics}\ }\textbf {\bibinfo
  {volume} {95}},\ \bibinfo {pages} {1274} (\bibinfo {year}
  {2004})}\BibitemShut {NoStop}%
\bibitem [{\citenamefont {M{\"u}ller}\ \emph {et~al.}(2013)\citenamefont
  {M{\"u}ller}, \citenamefont {Choi}, \citenamefont {Hellm{\"u}ller},
  \citenamefont {Ensslin}, \citenamefont {Ihn},\ and\ \citenamefont
  {Sch{\"o}n}}]{muller2013circuit}%
  \BibitemOpen
  \bibfield  {author} {\bibinfo {author} {\bibfnamefont {T.}~\bibnamefont
  {M{\"u}ller}}, \bibinfo {author} {\bibfnamefont {T.}~\bibnamefont {Choi}},
  \bibinfo {author} {\bibfnamefont {S.}~\bibnamefont {Hellm{\"u}ller}},
  \bibinfo {author} {\bibfnamefont {K.}~\bibnamefont {Ensslin}}, \bibinfo
  {author} {\bibfnamefont {T.}~\bibnamefont {Ihn}}, \ and\ \bibinfo {author}
  {\bibfnamefont {S.}~\bibnamefont {Sch{\"o}n}},\ }\href@noop {} {\bibfield
  {journal} {\bibinfo  {journal} {Review of Scientific Instruments}\ }\textbf
  {\bibinfo {volume} {84}},\ \bibinfo {pages} {083902} (\bibinfo {year}
  {2013})}\BibitemShut {NoStop}%
\bibitem [{\citenamefont {Aassime}\ \emph {et~al.}(2001)\citenamefont
  {Aassime}, \citenamefont {Johansson}, \citenamefont {Wendin}, \citenamefont
  {Schoelkopf},\ and\ \citenamefont {Delsing}}]{aassime2001radio}%
  \BibitemOpen
  \bibfield  {author} {\bibinfo {author} {\bibfnamefont {A.}~\bibnamefont
  {Aassime}}, \bibinfo {author} {\bibfnamefont {G.}~\bibnamefont {Johansson}},
  \bibinfo {author} {\bibfnamefont {G.}~\bibnamefont {Wendin}}, \bibinfo
  {author} {\bibfnamefont {R.}~\bibnamefont {Schoelkopf}}, \ and\ \bibinfo
  {author} {\bibfnamefont {P.}~\bibnamefont {Delsing}},\ }\href@noop {}
  {\bibfield  {journal} {\bibinfo  {journal} {Physical Review Letters}\
  }\textbf {\bibinfo {volume} {86}},\ \bibinfo {pages} {3376} (\bibinfo {year}
  {2001})}\BibitemShut {NoStop}%
\bibitem [{\citenamefont {Stehlik}\ \emph {et~al.}(2015)\citenamefont
  {Stehlik}, \citenamefont {Liu}, \citenamefont {Quintana}, \citenamefont
  {Eichler}, \citenamefont {Hartke},\ and\ \citenamefont
  {Petta}}]{StehlikPRapp2015}%
  \BibitemOpen
  \bibfield  {author} {\bibinfo {author} {\bibfnamefont {J.}~\bibnamefont
  {Stehlik}}, \bibinfo {author} {\bibfnamefont {Y.-Y.}\ \bibnamefont {Liu}},
  \bibinfo {author} {\bibfnamefont {C.~M.}\ \bibnamefont {Quintana}}, \bibinfo
  {author} {\bibfnamefont {C.}~\bibnamefont {Eichler}}, \bibinfo {author}
  {\bibfnamefont {T.~R.}\ \bibnamefont {Hartke}}, \ and\ \bibinfo {author}
  {\bibfnamefont {J.~R.}\ \bibnamefont {Petta}},\ }\href {\doibase
  10.1103/PhysRevApplied.4.014018} {\bibfield  {journal} {\bibinfo  {journal}
  {Phys. Rev. Applied}\ }\textbf {\bibinfo {volume} {4}},\ \bibinfo {pages}
  {014018} (\bibinfo {year} {2015})}\BibitemShut {NoStop}%
\bibitem [{\citenamefont {Maurand}\ \emph {et~al.}(2016)\citenamefont
  {Maurand}, \citenamefont {Jehl}, \citenamefont {Kotekar-Patil}, \citenamefont
  {Corna}, \citenamefont {Bohuslavskyi}, \citenamefont {Lavi{\'e}ville},
  \citenamefont {Hutin}, \citenamefont {Barraud}, \citenamefont {Vinet},
  \citenamefont {Sanquer}, \citenamefont {Franceschi},\ and\ \citenamefont
  {Jehl}}]{maurand2016cmos}%
  \BibitemOpen
  \bibfield  {author} {\bibinfo {author} {\bibfnamefont {R.}~\bibnamefont
  {Maurand}}, \bibinfo {author} {\bibfnamefont {X.}~\bibnamefont {Jehl}},
  \bibinfo {author} {\bibfnamefont {D.}~\bibnamefont {Kotekar-Patil}}, \bibinfo
  {author} {\bibfnamefont {A.}~\bibnamefont {Corna}}, \bibinfo {author}
  {\bibfnamefont {H.}~\bibnamefont {Bohuslavskyi}}, \bibinfo {author}
  {\bibfnamefont {R.}~\bibnamefont {Lavi{\'e}ville}}, \bibinfo {author}
  {\bibfnamefont {L.}~\bibnamefont {Hutin}}, \bibinfo {author} {\bibfnamefont
  {S.}~\bibnamefont {Barraud}}, \bibinfo {author} {\bibfnamefont
  {M.}~\bibnamefont {Vinet}}, \bibinfo {author} {\bibfnamefont
  {M.}~\bibnamefont {Sanquer}}, \bibinfo {author} {\bibfnamefont {S.~D.}\
  \bibnamefont {Franceschi}}, \ and\ \bibinfo {author} {\bibfnamefont
  {X.}~\bibnamefont {Jehl}},\ }\href@noop {} {\bibfield  {journal} {\bibinfo
  {journal} {Nat. Commun.}\ }\textbf {\bibinfo {volume} {7}},\ \bibinfo {pages}
  {13575} (\bibinfo {year} {2016})}\BibitemShut {NoStop}%
\bibitem [{\citenamefont {Betz}\ \emph {et~al.}(2015)\citenamefont {Betz},
  \citenamefont {Wacquez}, \citenamefont {Vinet}, \citenamefont {Jehl},
  \citenamefont {Saraiva}, \citenamefont {Sanquer}, \citenamefont {Ferguson},\
  and\ \citenamefont {Gonzalez-Zalba}}]{betz2015dispersively}%
  \BibitemOpen
  \bibfield  {author} {\bibinfo {author} {\bibfnamefont {A.}~\bibnamefont
  {Betz}}, \bibinfo {author} {\bibfnamefont {R.}~\bibnamefont {Wacquez}},
  \bibinfo {author} {\bibfnamefont {M.}~\bibnamefont {Vinet}}, \bibinfo
  {author} {\bibfnamefont {X.}~\bibnamefont {Jehl}}, \bibinfo {author}
  {\bibfnamefont {A.}~\bibnamefont {Saraiva}}, \bibinfo {author} {\bibfnamefont
  {M.}~\bibnamefont {Sanquer}}, \bibinfo {author} {\bibfnamefont
  {A.}~\bibnamefont {Ferguson}}, \ and\ \bibinfo {author} {\bibfnamefont
  {M.}~\bibnamefont {Gonzalez-Zalba}},\ }\href@noop {} {\bibfield  {journal}
  {\bibinfo  {journal} {Nano letters}\ }\textbf {\bibinfo {volume} {15}},\
  \bibinfo {pages} {4622} (\bibinfo {year} {2015})}\BibitemShut {NoStop}%
\bibitem [{\citenamefont {Betz}\ \emph {et~al.}(2016)\citenamefont {Betz},
  \citenamefont {Tagliaferri}, \citenamefont {Vinet}, \citenamefont
  {Brostr{\"o}m}, \citenamefont {Sanquer}, \citenamefont {Ferguson},\ and\
  \citenamefont {Gonzalez-Zalba}}]{betz2016reconfigurable}%
  \BibitemOpen
  \bibfield  {author} {\bibinfo {author} {\bibfnamefont {A.}~\bibnamefont
  {Betz}}, \bibinfo {author} {\bibfnamefont {M.}~\bibnamefont {Tagliaferri}},
  \bibinfo {author} {\bibfnamefont {M.}~\bibnamefont {Vinet}}, \bibinfo
  {author} {\bibfnamefont {M.}~\bibnamefont {Brostr{\"o}m}}, \bibinfo {author}
  {\bibfnamefont {M.}~\bibnamefont {Sanquer}}, \bibinfo {author} {\bibfnamefont
  {A.}~\bibnamefont {Ferguson}}, \ and\ \bibinfo {author} {\bibfnamefont
  {M.}~\bibnamefont {Gonzalez-Zalba}},\ }\href@noop {} {\bibfield  {journal}
  {\bibinfo  {journal} {Applied Physics Letters}\ }\textbf {\bibinfo {volume}
  {108}},\ \bibinfo {pages} {203108} (\bibinfo {year} {2016})}\BibitemShut
  {NoStop}%
\end{thebibliography}%
\end{document}